\documentclass[aps,prd,twocolumn,preprintnumbers,superscriptaddress,nofootinbib,floatfix]{revtex4-2}
\usepackage[utf8]{inputenc}
\usepackage{amssymb}
\usepackage{tensor}
\usepackage{xspace} 
\usepackage[pdftex]{hyperref}
\usepackage{hyperref}
\usepackage{graphicx}
\usepackage[caption=false]{subfig}
\usepackage{breakcites}
\usepackage{ragged2e}
\usepackage{amsmath}

\usepackage{setspace}

\hypersetup{
colorlinks,linkcolor=blue,citecolor=blue,urlcolor=blue
}

\usepackage{natbib}
\usepackage{tablefootnote}
\usepackage{footnote}
\usepackage{amsmath}
\usepackage{amssymb}
\usepackage[T1]{fontenc}

\newcommand{\SgrA}{SgrA$^{*}$}

\begin{document}

\title{The effect of environment in the timing of a pulsar orbiting  \SgrA}

\newcommand{\UNISA}{\affiliation{Dipartimento di Fisica ``E.R Caianiello'', Università degli Studi di Salerno,\\ Via Giovanni Paolo II, 132 - 84084 Fisciano (SA), Italy}}
\newcommand{\INFN}{\affiliation{Istituto Nazionale di Fisica Nucleare - Gruppo Collegato di Salerno,\\ Via Giovanni Paolo II, 132 - 84084 Fisciano (SA), Italy.}}
\newcommand{\INAF}{\affiliation{INAF, Osservatorio Astronomico di Cagliari, Via della Scienza 5, 09047 Selargius (CA), Italy}}
\newcommand{\ZARM}{\affiliation{ZARM, Universität Bremen, Am Fallturm 2, 28359 Bremen, Germany}}
\newcommand{\Biefi}{\affiliation{Fakul\"at f\"ur Physik, Universität Bielefeld, Postfach 100131, 33501 Bielefeld, Germany}}

\author{Amodio Carleo}
\email{acarleo@unisa.it}
\UNISA\INFN\INAF

\author{Bilel Ben-Salem}
\email{bilel.bensalem@zarm.uni-bremen.de}
\ZARM\Biefi

\date{\today}


\begin{abstract}
 Pulsars are rapidly rotating neutron stars emitting intense electromagnetic radiation that is detected on Earth as regular and precisely timed pulses. By exploiting their extreme regularity and comparing the real arrival times with a theoretical model (pulsar timing), it is possible to deduce many physical information, not only concerning the neutron star and its possible companion, but also the properties of the interstellar medium as well as tests of General Relativity. In light of recent works according to which dark energy may have an astrophysical origin, in this paper we investigate the effect of 'matter' on the propagation time delay of photons emitted by a pulsar orbiting a spinning black hole using the rotational Kiselev metric. We deduce an analytical formula for the time delay from geodesic equations, showing how it changes as the type of matter around the black hole (radiation, dust or dark energy) varies with respect to previous results, where  matter has not been taken into account.  It turns out that while  the spin $a$ only induces a shift in the phase of the maximum delay  without increasing or decreasing the delay, the effect of matter surrounding  the black hole results in a noticeable alteration of it. Our results show that dark energy  would give the strongest effect and that, interestingly, when the pulsar is positioned between the observer and the
black hole  a slightly lower pulse delay than in the
no-matter case appears. We estimated these effects for SGR J1745-2900, the closest magnetar orbiting SgrA*. 
\end{abstract}



\maketitle

\section{Introduction}
Pulsars are the result of the explosion of massive stars showing a repeated emission of radio waves which we detect as an extremely regular series of pulses. Since they are remarkably precise clocks, pulsars can be used to investigate many different aspects of physics, like testing theories of gravity, studying the magnetic field of the Galaxy and the interior of neutron stars, investigating the effect of the interstellar medium, and, last but not least, the detection of the gravitational waves background (GWB).   Most applications of pulsars involve a technique called “pulsar timing”, i.e. the measurement of the time of arrival (ToA)
of photons emitted by the pulsar, which are then compared with a theoretical model. With a collection of ToAs in hand, it becomes possible to fit a model of the pulsar’s timing behaviour, accounting for every rotation of the neutron star. Depending on whether the pulsar is isolated or binary, the multi-parameter fit gives  several important parameters (so-called ephemeris), like period, period derivative, orbital period (if binary), position in the sky, eccentricity, ecc. The first hint of the power of this method  was the case of the binary system PSR B1913+16, whose orbital decay  agree with their predicted values to better than 0.5\% \citet{taylor1989New}: the observed accumulated shift of periastron is  in excellent agreement with the General theory of Relativity, leaving little room for alternative theories of gravity. Over the years there have been numerous studies on possible violations of General Relativity (GR) using the timing of the pulsars  \citep[e.g.][]{Stairs_2003,https://doi.org/10.48550/arxiv.2204.13468}, revealing that pulsars are a great research tool in this field, offering also the possibility  to test the no-hair theorem as well as the
cosmic censorship conjecture \citep{Liu_2012,Izmailov:2019cqr}. \\
More recently, pulsar timing is also used to detect and characterize the low-frequency gravitational wave universe through timing an array of approximately 100 millisecond pulsars (MSP) using the largest radio telescopes in the world. Indeed,  gravitational waves will cause changes in the travel times of pulses between pulsars and the Earth, detectable as perturbations in pulsar time-of-arrival measurements. Most importantly, this signature will show a characteristic sky correlation, predicted by Einstein’s theory of General Relativity, detectable by correlating the data from all of the pulsars in the array. Any unmodelled effects will appear in the timing residuals, and the timing model is revised and/or extended accordingly: trends in the residuals are indicative of non-optimized parameters, while white noise residuals suggest a good timing model. Generally, the timing models just requires the precise modelling of the pulsar’s rotation, orbital motion and the signal’s propagation in space, and not the details of the radiation’s physics or emission
mechanism. In the case of GWB, the unmodelled residulas would be a red noise at nHz frequencies \citep{detweiler}. Even if to date there has not been a reveal of such background,  the road looks promising \citep{Goncharov_2021,ipta}. \\
Usually, the timing model to predict ToAs is based on Damour and Deruelle’s
approach using a post-Newtonian expansion to treat the relativistic
two body problem \citep{deruelle}, and its corresponding
relativistic effects are described by a set of post-Keplerian parameters, see e.g. \citep{2006MNRAS.372.1549E}. However, the validity of the post-Newtonian approximation (that assumes a weak
field) it is no longer guaranteed for a pulsar orbiting closely a supermassive black hole (SMBH), in particular if  
 pulsar, black hole and observer are (nearly) aligned. When the weak field approximation holds, then three \textit{distinct} delay terms arise: the Roemer delay, the Shapiro delay and the Einstein delay. The Roemer delay  is the  difference of arrival time between the case in which the Earth is “on the same side” of the pulsar and
the one in which is on the other side during the orbit around the sun. In order to avoid the modulation induced by Earth's orbit, ToAs are referred to the Solar System barycenter (SSB), where coordinate time is defined as $t_{SSB} = t_{em} + (1/c)|\mathbf{r}_{p} -\mathbf{r}_{b}|$, where $t_{em}$ is the time of photon emission, and $\mathbf{r}_{p,b}$ is the position of the pulsar or SSB (usually calculated using distant quasars).  The Shapiro delay is the (always positive) additional delay to take into account the deviation of light caused by the gravitational
field of the Solar System  and it is easily obtained by solving the geodetic equations in the weak field approximation, as the case of the Solar System. Finally, the Einstein delay is due to the difference between coordinate time $t$ of the pulses and the proper time $\tau$ of observation, which is different because of a relative motion w.r.t. the pulsar. In the case of a binary pulsar, there will be a Roemer, Shapiro and Einstein delays  also for the binary system \footnote{In the case of a binary pulsar, the time of emission is localized at the binary barycenter.}, with the difference that, in
the case of a binary pulsar, general relativistic effects are much more important than in the Solar System,
being a relativistic system. This implies that the expressions for the above three different delays  become more complex, since we  have to treat a full general relativistic two-body problem. The two-body problem does not have an exact solution in GR and hence solving it requires a perturbative post-Newtonian expansion, whose coefficients (post-Keplerian parameters) can be inferred  by pulsar timing. They are linked not only to the shape of the orbit (like eccentricity and semi-major axis), but also to other non-Newtonian effects, such as the above mentioned Shapiro and Einstein delays or the decay of the orbit due to gravitational radiation labeled with $\dot{P}$. First post-newtonian order terms  are usually sufficient, but if data set is accurate, higher order contributions can in principle be added in order to get even more information on the binary system (for a review see e.g. \citep{Manchester_2015}).      \\
Among the various non-Keplerian parameters, recently the effect of dragging was studied: a compact companion induce a delay in the ToAs due to the frame dragging on the pulsar's radiation, if the latter rotates close enough. In \citep{Bilel}, a comparison between an exact analytical result for the frame dragging time delay and two post-Newtonian derivations was analysed. The exact formula was derived as the difference between exact geodesic solution for a Kerr black hole (the compact companion) and the equivalent solution for a Schawarzchild \citep{Eva_2019}. The conclusion is that post-Newtonian based treatments overestimate the frame dragging effect on the lightlike signals, in particular around and after superior conjunction, hence the analytical solution provide a more reliable and accurate approach, especially in extreme mass ratio binary configurations, as the case of a pulsar orbiting Sagittarius A$^{*}$ (SgrA$^{*}$).  In this type of setup, where the pulsar is considered as a test particle, the effect of the spin is not negligible and analytical formulas for the propagation delay are necessary, since fully non-linear numerical approaches suggest that easier post-Newtonian treatments may not be very accurate \citep{Zhang_2017,Kimpson_2019}.  \\
In the wake of these results and since recently evidence has emerged that dark energy  could have an astrophysical origin  \citep{BH_DE_inside,BH_DE_inside_2}, in this paper we study the further effect of the presence of matter (radiation, dust or dark energy) in the surrounding of the central massive black hole. The aim of this work is to investigate how the  propagation time delay is affected by the type of matter, whose presence is certainly not negligible in the case of SgrA$^{*}$. This could serve, in principle, to understand the environment in which the binary system is located by exploiting the timing of the pulsars, provided that this effect is separable from the others.  The outline of the paper is as follow. In Section 2 we derive  the equations of motion for lightlike geodesic using a rotational Kiselev metric, and characterize the corresponding parameter space $(\lambda,q)$ where $\lambda$ is the adimensional angular momentum and $q$ is the Carter constant. In Section 3 we analytically  solve them  using the Mino time and the elliptic functionals, while  in Section 4 we apply the   results to three different type of matter (radiation, dust and dark energy; see later) and compare them to simpler Kerr case. Finally, we close the manuscript with a summary and an outlook in Section 5.

\section{Geodesic Equations}

\subsection{The Kiselev metric}

\begin{figure}
\includegraphics[width=\columnwidth]{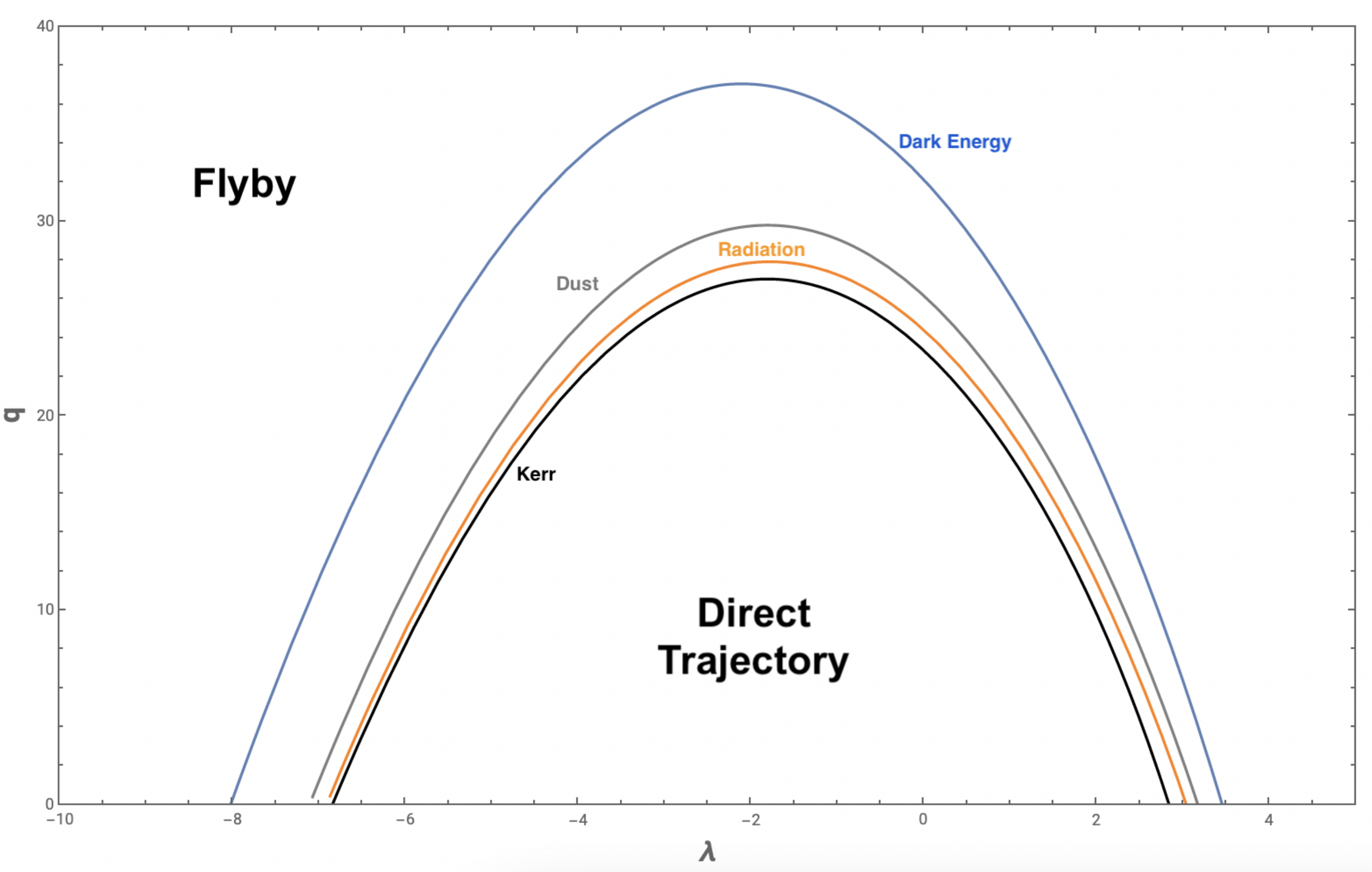}
   \caption{Curves, in the parameter space ($\lambda$,$q$), of double roots \textit{outside} of the horizons for the radial potential $R(r)$ in the case of radiation (orange), dust (gray) and dark energy (blue) orbiting a central black hole, when $a=0.9$ and $c=0.1$ ($M=1$). Kerr black hole (black) with no matter is also shown.  When $q=0$, the corresponding $\lambda$ values, $\lambda_{\pm}$, on the double roots curve, coincide with the impact parameter of the unstable circular photon orbits. Above each curve, photons enter from infinity, reach a turning point \textit{outside} of the horizons and returns to infinity (flyby); below each curve, photons move directly between infinity and the horizon,  with no turning points (direct trajectory). The latter behavior also occurs when $q<0$ (not shown in figure). }
   \label{fig:1}
\end{figure}

The solution of Einstein's field equation for a Schwarzchild black hole surrounded by quintessence (a type of dark energy) has been obtained in Ref.~\citep{Kiselev:2003,Ghosh}. Even if in these works only quintessence is considered, the Kiselev solution contemplates any type of energy-matter, once a state parameter has been established. Indeed, a rotational Kiselev black hole looks like \citep{Toshmatov:2015npp}. \begin{equation}\label{eq1}
\begin{array}{ccc}
d s^{2}=-\Big(1  -\frac{2 M r+ c r^{1-3 \omega}}{\Sigma^{2}}\Big) d t^{2}  +\frac{\Sigma^{2}}{\Delta} d r^{2} \\[6pt] -\frac{2 a \sin^{2} \theta\left(2 M r+ c r^{1-3 \omega}\right)}{\Sigma^{2}} d \varphi d t 
+\Sigma^{2} d \theta^{2} \\[6pt]
    +\sin^{2} \theta\left(r^{2}+a^{2}+a^{2} \sin^{2} \theta \frac{2 M r+ c r^{1-3 \omega}}{\Sigma^{2}}\right) d \varphi^{2}
    
\end{array}
\end{equation}
where we defined
\begin{equation*}
 \Delta=r^{2} -2Mr +a^{2}-cr^{1-3w}, \; \; \;  \Sigma^{2}=r^{2}+a^{2}\cos^2{\theta} . \end{equation*}
$M$ is the mass of the black hole and $a$ is the spin parameter. Moreover, $c$ is the strength parameter and $w$ defines the EoS, $p=w\rho$. Eq.~(\ref{eq1}) is the rotational symmetry solution for a black hole wrapped in any kind of energy-matter definable by the EoS. In general, for dark energy, we would expect $w<0$. In the following, we will investigate dust ($w=0$) and radiation ($w=1/3$), as well as a dark energy like component with $w=-1/3$  \citep{Melia:2014vva}. \\
The number of horizons depends on the value of $w$. For $-1\leq w <-1/3$, $\Delta=0$ has three positive solutions, corresponding to a Cauchy horizon, an event horizon and a cosmological horizon. For $w=\pm1/3$ and $w=0$, the cosmological horizon disappears and only two horizons, $r_{\pm}$, exist. As an upper limit on the strength parameter $c$ for the cases we want to study, we choose $c<1$, compatibly with \citep{carleo1}. We decided not to make a perturbation for small values $c\ll1$, since the presence of matter around a supermassive black hole might not be negligible, as also shown by the image of the SMBH in the center of M87$^{*}$ \citep{M87}. Notice, finally, that the drag effect is higher at higher $|w|$ values, and, when $w$ is fixed, is stronger at higher $c$ values; when $c\ll 1$, the dependence on $c$ is approximately linear. Therefore, these effect could combine with the non-zero spin parameter $a$ and create a degeneracy between  different parameters. \\

\subsection{Geodesic Equations}
Using the metric components $g_{\mu\nu}$ from the line element $ds$,  geodesic equations are given by 
\begin{equation}\label{eq2}
    g_{\mu \nu} \dfrac{dx^{\mu}}{d\lambda}\dfrac{dx^{\nu}}{d\lambda}=\epsilon
\end{equation}
where $\lambda$ is an affine parameter and $\epsilon=0,-1$. While in Schawarzchild spacetime orbits starting in the plane (for example $\theta=\pi/2$) remain planar, in Kerr and hence Kiselev metric this is not true and an additional motion constant is needed. Here, to find equations of motion (EOM), we adopt the well-known Hamilton-Jacobi equation and Carter constant separable method \footnote{A similiar treatment has been  done for a specific state parameter value ($w=-2/3$) in \cite{2017}. } \citep{carter}.  Therefore, action $S$ must satisfy:  
\begin{equation}\label{action}
    \dfrac{\partial S}{\partial \lambda} = \dfrac{1}{2}k\lambda -E t + L \varphi + S_{r}(r)+S_{\theta}(\theta)
\end{equation}
where $p_{\alpha}\doteq \partial S / \partial x^{\alpha}$, $L\doteq p_{\varphi}$, $E\doteq - p_{t}$,  $k=0$ for photons and $k=-m^{2}$ for massive particles. With the inverse metric components 
\begin{equation}\label{inverse}
g^{\mu \nu}=\left(\begin{array}{cccc}
g^{tt} & 0 & 0 & - \dfrac{a(2Mr+cr^{1-3w})}{\Sigma \Delta} \\
0 & \dfrac{\Delta}{\Sigma} &  0 & 0 \\
0& 0 & \dfrac{1}{\Sigma} &  0 \\
 - \dfrac{a(2Mr+cr^{1-3w})}{\Sigma \Delta}& 0 & 0 & g^{\varphi\varphi} 
\end{array}\right)
\end{equation}
where 
\begin{equation}
\begin{array}{ll}
  g^{tt} =  \dfrac{a^{2}\Delta\sin^2{\theta}-(a^{2}+r^{2})^2}{\Delta \Sigma }   \\
    g^{\varphi \varphi} =  \dfrac{\Delta-a^{2}}{\sin^2{\theta}}{\Sigma\Delta\sin^2{\theta}} .
\end{array}
\end{equation}
After rearranging, Eq. (\ref{action}) comes down to two separated equations, namely
\begin{equation}
    \begin{array}{lll}
         E^{2}a^{2}{\cos^2{\theta}}+ka^{2}{\cos^2{\theta}} - \left( \dfrac{\partial S}{\partial \theta} \right)^{2} - L^{2}{\cot^2{\theta}}  = \mathcal{C}\\
         kr^{2}-(aE-L)^{2}-\Delta\left(\dfrac{\partial S}{\partial r}\right)^{2}+ \dfrac{1}{\Delta}\left[ E(r^2 +a^2) -aL \right]^{2} = \mathcal{C} 
    \end{array}
\end{equation}
where $\mathcal{C}$ is a separation constant. From definitions of energy $E$ and angular momentum $L$, the following relations hold
\begin{equation}
\begin{array}{ll}
\dot{t}=-\frac{E}{g_{tt}}\left[ 1 + g_{t\varphi} \left(\lambda + \frac{g_{t\varphi} }{g_{tt}}\right) \left(g_{\varphi\varphi}- \frac{g_{t\varphi}^{2}}{g_{tt}}\right)^{-1}\right] \\
\dot{\varphi}=E \left(\lambda + \frac{g_{t\varphi}}{g_{tt}}  \right) \left(g_{\varphi\varphi}- \frac{g_{t\varphi}^{2}}{g_{tt}}\right)^{-1} \,\ ,
\end{array}
\end{equation}
where $\lambda \doteq L/E$. Hence, the geodesic equations are 
\begin{equation}\label{eq:t}
    \dot{t} = \dfrac{(r^2+a^2)(r^2+a^2-a\lambda)}{\Delta}-a(a-\lambda)+a^2\cos^2{\theta}
\end{equation}
\begin{equation}\label{eq:varphi}
   \dot{\varphi} = \dfrac{a(r^2+a^2-a\lambda)}{\Delta}-a+\dfrac{\lambda}{\sin^2{\theta}}
\end{equation}
\begin{equation}\label{eq:teta}
   \dot{\theta}^2 = q + \cos^2{\theta} \left[ \left(1 + \dfrac{k}{E^2}\right)a^2 -\dfrac{\lambda^{2}}{\sin^2{\theta}} \right] =  \Theta(\theta)  
\end{equation}
\begin{equation}\label{eq:r}
     \dot{r}^2 = -\Delta\left[q-\dfrac{k}{E^2}r^2+(\lambda -a)^2 \right] + \left( r^2+a^2-\lambda a \right)^2 = R(r) .
\end{equation}
Here, a dot means derivative w.r.t. the so called Mino time $\gamma$, which satisfies the condition $dx^{\mu} = (\Sigma/E) p^{\mu} d\gamma$, while $q\doteq  \mathcal{C}/E^{2}$. Quantities $m$, $E$, $L$, $\mathcal{C}$ are  constants of motion for the EOMs (\ref{eq:t})-(\ref{eq:r}).   Notice that when $c=0$ Eqs.(\ref{eq:t})-(\ref{eq:r}) are equal to the Kerr case \citep{Bilel}, 
 although they formally remain similar even when $c$ is non-zero. Furthermore, as for a Kerr metric, $\Theta(\pi/2) = 0 $ $\longleftrightarrow$ $q=0$, i.e. a geodetic lies entirely in the equatorial plane if and only if $q=0$. One difference is the passage for a point with $r=0$ and $\theta \not= \pi/2$ when $w=1/3$. Indeed, the condition $R(0) \ge 0$ for $w=1/3$ implies 
 \begin{equation}
     q\le \dfrac{c (a-\lambda)^2}{(a^2-c)} \ge 0
 \end{equation}
where we assumed $c<<1$. In this regime, $r=0$ is approachable for both negative and positive values of $q$, while for $w \in \{-1,-2/3,-1/3,0 \}$ this happens only for negative values ($q<0$), as in Kerr metric. On the other hand, from the positivity condition $\Theta(\theta) \ge 0$, motion is allowed only when \citep{Gralla_2020}
 \begin{equation}\label{eq14}
     q \geq\left\{\begin{array}{ll} 0 & |\lambda| \geq a \\ -(|\lambda|-a)^2 & |\lambda| \leq a \end{array}\right.
 \end{equation}
which border an allowed region for the parameter space $(q,\lambda)$ and is independent of the new parameter $c$.  

\subsection{Roots of radial potential}
The calculation and classification of the roots of the radial potential $R(r)$ proceed in a similar way to the Kerr case \citep{Gralla_2020}. We consider here the most interesting cases for our scopes, i.e. $w \in \{\pm 1/3,0\}$. \\
When $w=-1/3$, the solution of equation $R(r)=0$ is given by
\begin{equation}\label{eq15}
\begin{aligned} r_1 & =-z-\sqrt{-\frac{A}{2}-z^2+\frac{B}{4} z^2} \\ r_2 & =-z+\sqrt{-\frac{A}{2}-z^2+\frac{B}{4}} \\ r_3 & =z-\sqrt{-\frac{A}{2}-z^2-\frac{B}{4 z}} \\ r_4 & =z+\sqrt{-\frac{A}{2}-z^2-\frac{B}{4 z}} \end{aligned}
\end{equation}
where 
\begin{equation*}
    \begin{aligned}
        A & = a^2 -q -\lambda^2 + c\chi , \; \; \; \; \; 
        B  = 2M \chi \\
        C & = -a^2 q , \; \; \; \; \;\; \; \; \; \; \; \; \; \; \; \; \; \; \; \; \; \; 
        \chi  = q + (\lambda -a)^2 \ge 0.
          \end{aligned}
\end{equation*}
A very similarly computation is required for the remaining two values of $w$ that we consider, hence we omit it for brevity.  As we expect, the discrepancy between the Kerr and Kiselev radial potential roots increases as the strength  parameter $c$ increases. The effect is more evident for negative values of $w$. When $w=0$, the additional term appearing always competes with the mass term $M r$, being therefore almost negligible. \\
In order to find quadruple roots of the potential, we impose the form $R(r) = (r-r_{0})^4$ and after comparing with its explicit expression, we obtain 
\begin{equation}
    q=0 , \; \; \; \; \lambda = a 
\end{equation}
besides that $r_{0}=0$. Here, we assumed $a\not=0$. Triple solutions satisfy the relations $R(r)=R'(r)=R''(r)=0$, whose solution is
\begin{equation}
    q=\dfrac{8r^3}{2M} -(\lambda-a)^2, \; \; \; \;  \lambda = \dfrac{M(a^2+3r^2)-2r^3(1-c)}{aM}
    \end{equation}
and hence
\begin{equation}
    r= \dfrac{M-\left[M(M^{2}-(1-c)a^2)\right]^{1/3}}{(1-c)}.
\end{equation}
Finally, double solutions ($R'(r)=R(r)=0$) occur when
\begin{equation}
\begin{aligned}
    q &=\dfrac{r^3}{a^2} \left[ \dfrac{4M\Delta}{\left( M-r(1-c) \right)^2} - r  \right], \\
    \lambda &= a + \dfrac{r}{a}\left[  r- \dfrac{2\Delta}{r(1-c)-M}\right].
\end{aligned}
\end{equation}
We then evaluate these relations on the border of the admissible region (\ref{eq14}), i.e. $q=0$ and $q= -(\lambda\pm a)^2$. In the first case, three real solutions for $r$ are possible:
\begin{equation}\label{eq:20}
    r= e \cos \left[ \dfrac{1}{3} \arccos \left(  \dfrac{2a^2(1-c)-M^2}{M^2}  \right) +\dfrac{2\pi k}{3}\right] + e
\end{equation}
with $k=0,1,2$ and $e=2M/(1-c)$. When $q= -(\lambda + a)^2$, four real and distinct solutions for $r$ are allowed; their structure is similar to (\ref{eq15}) and  we omit them for brevity. Finally, in the last case, $q= -(\lambda - a)^2$,  in addition to the Kerr case analogous solution, i.e. $r_{\pm}$, other options appear, namely $r_{N}= \pm a/\sqrt{c}$. Since double solution curves separates the parametric $(q,\lambda)$ regions with two or four real solutions, once the values of mass $M$, spin $a$ and strength of matter $c$ have been fixed,  different behaviours of the roots are delimited by such curves, which, indeed, border the range of radial coordinate $r$ for which (\ref{eq14}) holds, i.e. motion is allowed. In particular, when $q=0$, the corresponding $\lambda$ values on the double roots curve, $\lambda_{\pm}$, coincide with the impact parameters of the so-called  \textit{unstable circular photon orbits} (counter-rotating and co-rotating). Radii of such (equatorial) orbits are simply given by the largest roots (\ref{eq:20}), and in the case $w=-1/3$ are
\begin{equation*}
    r_{ph}^{\pm} = e \cos \left[ \dfrac{1}{3}\arccos \left( \dfrac{2a^2(1-c)-M^2}{M^2} \right) + k_{\pm}\dfrac{4 \pi}{3} \right] +e 
\end{equation*}
where $k_{+}=0$, $k_{-}=2$. When $c=0$, they reduce to the well-known Kerr results.  Notice that with this notation $\lambda_{+}\equiv \lambda (r_{ph}^{+} )$ and $\lambda_{-}\equiv \lambda(r_{ph}^{-})$, it will be $\lambda_{+}<\lambda_{-}$. A possible trend for the three different types of matter is showed in Fig. (\ref{fig:1}), where each point of the curves corresponds to a choice  of parameters ($\lambda$,$q$) for which two  roots for the potential $R(r)$ coincide ($r_{3}=r_{4} > r_{+}$).    Above the  curve, four real roots of the potential are allowed, two of which outside of the horizons. Conversely, below the curve ($\lambda_{+} <\lambda < \lambda_{-}$ when $q=0$) no real roots outside of the horizons exist. Since $R(r)>0$ at infinity, radial potential is positive in the ranges $r<r_{1}$, $r_{2}<r<r_{3}$ and $r>r_{4}$. Therefore, only in the first case   photons enter from infinity, reach a turning
point \textit{outside} of the horizons (at $r = r_{4}$), and returns to infinity \footnote{In addition to this motion (flyby), for parameter space points above the curve a bound orbit is  also possible, but that is not interesting for our purposes.}; in the second case, photons move directly between infinity and the horizon (there may be turning points \textit{inside} the horizon). The presence of matter also reduces the parametric region corresponding to four real roots all inside the horizons (not shown in Fig. (\ref{fig:1})). As we expect, the effect is more evident for larger value of $c$, but it is more pronounced in the presence of radiation rather than dark energy. \\
In what follows, we will deal with a beam of photons from a pulsar orbiting a black hole in the equatorial plane. In this case, the trajectory followed can be of only two types: either the photons move directly from the pulsar to the observer, or they first encounter a turning point outside of the horizons ($r_{4}$) and then move from it to the observer. The parametric region corresponding to the first case (direct trajectory)  is made up of all points below the curve (even for $q<0$), while the second case (flyby) describes trajectories for photons with motion parameters above the curve. Overall, the presence of matter attached to a rotating black hole increases the region of parameter space in which a direct trajectory towards a distant observer is possible, at the expense of a flyby. More precisely, it is easy to find out the behaviour of  the emitted photons from considerations on the potential shape and its positivity ranges \cite{Gralla_2020}. \\
Computation of quadruple, triple, and double roots for the remaining two cases $w=0$, $w=1/3$ are reported in Appendix A.

\section{Time Delay}

In this section we analytically solve Eqs. (\ref{eq:t})-(\ref{eq:r}) using elliptic integrals in the Legendre form, following the strategy used in \cite{Bilel}. As in the previous section, for brevity we show here only the case $w=-1/3$. Final results for different cases are however summarized in Appendix B. \\
\subsection{The exact time delay}

Combining Eqs. (\ref{eq:t}), (\ref{eq:teta}) and (\ref{eq:r}), one get an equation for $t$ in integral form, i.e.
\begin{equation}\label{eq20}
    \mathrm{c}(t_{a}-t_{e}) = \int_{\gamma_{r}} \dfrac{G(r)}{\Delta \sqrt{R(r)}} dr + \int_{\gamma_{\theta}} \dfrac{a^2 \cos^2{\theta}}{\sqrt{\Theta (\theta)}} d\theta
\end{equation}
where $\mathrm{c}$ is the speed of light (not to be confused with $c$) and we have defined
\begin{equation}\label{eq21}
    G(r) = r^2 (r^2+a^2+ac(a-\lambda)) + 2Mra(a-\lambda).
\end{equation}

We notice that the term proportional to $c$ is  the novelty with respect to the Kerr case (as well as the different definition of $\Delta$). On the other hand, the angular integral remains unchanged. The integral path $\gamma_{r}$ starts at the radial point of emission $r_e$ and either runs directly to infinity (direct trajectory) or first decreases in radius towards a turning point outside of the horizons ($r_4$) and then return to infinity (flyby), according to  the motion parameters of photons (see Fig. (\ref{fig:1})). Therefore, we split radial integral as
\begin{equation} \label{eq22}
    \left( \int_{r_4}^{\infty}  \pm \int_{r_4}^{r_e}\right) \dfrac{G(r)}{\Delta \sqrt{R(r)}} dr
\end{equation}
where we choose the minus sign for a direct trajectory \footnote{Notice that, in this case, $r_{4}$ is just a generic point between the emission  and the observer position, which we fixed at infinity, i.e. $r_a=\infty$. Therefore $r_e<r_4<\infty$.} and the plus sign for a flyby motion.  Similarly, the angular integral in Eq. (\ref{eq20}) can be written as \citep{Bilel}
\begin{equation}
    \left( \int_{0}^{u_e}  \pm \int_{0}^{u_a}\right) \dfrac{a^2 u}{2  \sqrt{U(u)}} du
\end{equation}
where we defined
\begin{equation*}
    U(u)= u(q+u(a^2-\lambda^2-q)-a^2u^2)
\end{equation*}
and we changed the integration variable to $u=\cos^2\theta$. Hence, $u_e$ and $u_a$ represent the emission and the observer latitudinal positions, respectively. Here, we choose the plus sign if the equatorial plane is crossed and the negative sign else \footnote{We adopt the usual convention that $\theta=0$ corresponds to the north pole. In this way, $\theta=\pi/2$ correspond to the equatorial plane.}. The above equation strictly only holds in absence of latitudinal turning points $u_{\pm}$ (the non-zero roots of $U(u)$). If latitudinal turning points are encountered, we have to add complete integrals in the form $\int_{0}^{u_{+}}$ if $q>0$ or   $\int_{u_{-}}^{u_{+}}$ if $q<0$. However, below we will assume that we are not in these cases. \\
The integral (\ref{eq22}) can then be solved exactly in terms of elliptic integrals. We notice that similar expressions for the time evolution of lightlike geodesics in Kerr metric have been derived before, in slightly different ways than the one used here and in \cite{Bilel}. For example, \cite{Dexter_2009} give an expression using partly Carlson's elliptic integrals. Expressions in terms of Weierstrass functions have been derived in \cite{Eva_2019}, while in \cite{Gralla_2020} Jacobi elliptic integrals in 'Jacobi form' (instead of Legendre form employed here) were used. Here, we are not concerned with the optimal choice; what is certain is that having an analytical solution allows to avoid  divergences which would be impossible to avert in a purely numerical calculation. The result is (details in Appendix B)
\begin{equation}\label{eq24}
\arraycolsep=1.4pt\def\arraystretch{2.2}
\begin{array}{ccc}
\mathrm{c}(t_{a}-t_{e}) =&  T_{r}(\infty,\lambda_e,q_e) \pm T_{r} (r_e,\lambda_e,q_e) \\
&+ |T_{u}(u_e,\lambda_e,q_e) \pm T_{u} (u_a,\lambda_e,q_e) |
\end{array}
\end{equation}
 
with the definitions
\begin{equation}\label{eq25}
\arraycolsep=1.8pt\def\arraystretch{2.6}
\begin{array}{l}
T_{u}(u, \lambda, q)=\int_0^u \frac{a^2}{2} \frac{u d u}{\sqrt{U_{\lambda, q}}} \\
=\frac{a}{\sqrt{u_{+}-u_{-}}}\left[u_{-} F(v, w)+\left(u_{+}-u_{-}\right) E(v, w)-\frac{u_{+} v \sqrt{\mathrm{1}-v^2}}{\sqrt{1-w^2 v^2}}\right]
\end{array}
\end{equation}
and   
\begin{equation}\label{eq-26}
\arraycolsep=1.8pt\def\arraystretch{2.6}
\begin{array}{l}
T_{r}(r, \lambda,q;c)=\delta \cdot \Big[F(x, k) \cdot \Big(4 M^2 \gamma^{-3}-a^2 c^2 \gamma^{-2}-a c \lambda \gamma^{-1}\\
+2 M r_3 \gamma^{-2} 
+\frac{1}{2} \gamma^{-1} \cdot\big[r_1\left(r_3-r_4\right)+r_3\left(r_3+r_4\right)\big]+\frac{B_+ l}{l_+}+\frac{B_- l}{l_-}\Big) \\
+E\big(x, k\big) \cdot\Big(-\frac{1}{2} \gamma^{-1} \cdot\big(r_4-r_2\big)\big(r_3-r_1\big)\Big) \\
+\Pi(x, l, k) \cdot\Big(2 M r_4 \gamma^{-2}-2 M r_3 \gamma^{-2}\Big) \\
+\Pi(x,l_{+}, k) \cdot\Big(B_{+}-\frac{l B_{+}}{l_{+}}\Big)+\Pi\big(x, l_{-}, k\big) \cdot\Big(B_{-}-\frac{l B_{-}}{l_{-}}\Big)\Big] \\
+\frac{\sqrt{R(r)}}{r-r_3} \gamma^{-1} \\
\end{array}
\end{equation}
where $r_{1,..,4}$ are the radial potential roots (\ref{eq14}), $u_{\pm}$ the non-zero roots of $U,$ $r_{\pm}$ are the horizons, and for brevity, we defined (for $v$, $w$, $x$, $k$ and $B_{\pm}$ see Appendix B) 
\begin{equation*}
    \delta = \frac{2}{\sqrt{\left(r_4-r_2\right)\left(r_3-r_1\right)}} , \; \; \; \; \; \; \; \gamma = 1-c ,
\end{equation*}
\begin{equation*}
   l = \dfrac{r_1-r_4}{r_1-r_3} , \; \; \; \; \; \; \; l_{\pm} = \dfrac{l (r_3-r_{\pm})}{r_4-r_{\pm}}. 
\end{equation*}
The functions $F$, $E$ and $\Pi$ appearing in Eq.  (\ref{eq-26}) are the well-known elliptic functions of first, second and third kind, respectively (see again Appendix B). To obtain Eq. (\ref{eq-26}) we also used the vanishing rule $\sum_{i=1}^{4} r_{i} = 0$. The above equations reduce to Kerr ones \citep{Bilel} when $c\rightarrow 0$. As in Kerr, some divergences appear in Eq.(\ref{eq-26}), specifically in $\Pi(x,l,k)$ and in the last term, which is not clear how to handle in fully numerical calculations. In Eq. (\ref{eq24}), the constants of motion $\lambda$ and $q$ only  depend on the emission point, since we have fixed the arrival point at infinity. Furthermore, in the case of equatorial orbits ($\theta=\pi/2$) the angular integral in Eq. (\ref{eq20}), and hence $T_{u}$ in Eq. (\ref{eq25}), are vanishing. \\
To avoid divergences, it is usual to subtract the time delay w.r.t. a fixed reference point from  the actual time delay (\ref{eq24}), namely
\begin{equation}\label{eq27}
\arraycolsep=1.6pt\def\arraystretch{1.9}
 \begin{array}{l}
\Delta t_{e x}\left(t_e, \varphi_e, u_e;c\right)=  \left(t_a-t_e\right)-\left(t_a-t_{\text {ref }}\right) \\
=  \frac{1}{\mathrm{c}}\Big[T_r\left(\infty, \lambda_e, q_e\right) \pm T_r\left(r_e, \lambda_e, q_e\right) \\
 +|T_u\left(u_e, \lambda_e, q_e\right) \pm T_u\left(u_a, \lambda_e, q_e\right)|\Big] \\
 -\frac{1}{\mathrm{c}}\Big[T_r\left(\infty, \lambda_{\text {ref }}, q_{\text {ref }}\right) \pm T_r(r_{\text {ref }}, \lambda_{\text {ref }}, q_{\text {ref }}) \\
 +|T_u\left(u_{\text {ref }}, \lambda_{\text {ref }}, q_{\text {ref }}\right) \pm T_u\left(u_a, \lambda_{\text {ref }}, q_{\text {ref }}\right)|\Big],
\end{array}   
\end{equation}
where $\lambda_{ref},q_{ref}$ are the angular momentum and Carter constants at the reference point and $c$ in the r.h.s. means that the expression is evaluated  in presence of matter \footnote{In the following, we simply denote Eq. (\ref{eq27}) with $\Delta t_{ex}(c)$. This implies that $\Delta t_{ex}(c=0)$  reproduce same results of \cite{Bilel}, i.e. in absence of surrounding matter.}. The ascending node w.r.t. the plane of the sky ($\phi=-w$) is used as the reference point, since the time delay is zero for photons leaving the pulsar in such a position. Actually, the Roemer, first-order Shapiro (when $e=0$) and geometric delays all vanish at the ascending node ($\phi=-w$), but other delays, like Einstein delay, second-order Shapiro delay, as well as the exact formula (\ref{eq27}) do not vanish at the ascending node, and the addition of individual offset to each last  types of delay is usually needed to have a vanishing delay point.   
\subsection{Orbital parameters}

\begin{figure}
\centering
\includegraphics[width=0.4\textwidth]{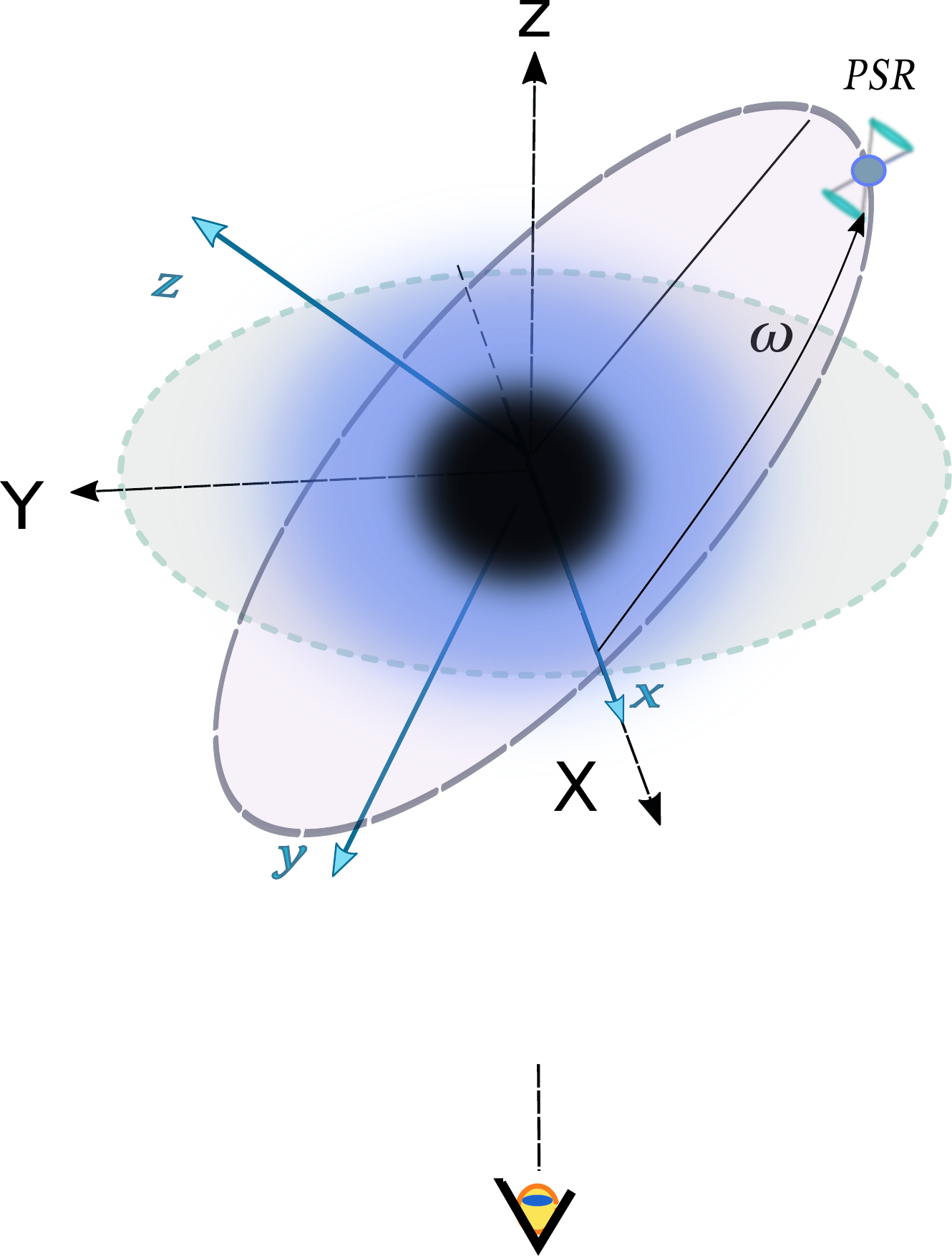}
\caption{Orientation of the black hole-pulsar system in the sky with respect to an observer sitting at infinity. The blue shaded region around the black hole reflects the possible environment of dark energy, radiation or dust investigated in this work.}
\label{fig:diagram1}
\end{figure}

In order to fully calculate Eq. (\ref{eq24}), we need the coordinates of the emission point ($r_e,\theta_e,\varphi_e,t_e$) on the pulsar orbit as well as the observer latitude $\theta_a$ ($r_a=\infty$ and we assume, for simplicity, $\varphi_{a}=0$). The position of the emission point follows the pulsar's orbit around the black hole (see Fig. (\ref{fig:diagram1})); for each point of the orbit, we need the motion parameters ($\lambda_e,q_e$) of geodesics connecting to points to infinity. As pointed out by \cite{Bilel}, there is no general analytical solution to such problem (emitter-observer problem).
To obtain $\lambda$ and $q$, in general, one needs to numerically solve both the equations 
\begin{equation}\label{eq28}
    \int_{\gamma_r} \dfrac{dr}{\sqrt{R(r)}} = \int_{\gamma_{\theta}}\dfrac{d\theta}{\sqrt{\Theta(\theta)}} ,
\end{equation}
\begin{equation}\label{eq29}
    \varphi_a-\varphi_e = \int_{r}\dfrac{2Mra-a^2\lambda+acr^2}{\Delta \sqrt{R(r)}} dr + \int_{\gamma_{\theta}} \dfrac{\lambda}{\sin^2(\theta) \sqrt{\Theta(\theta)}} d\theta
\end{equation}
where the first integral in Eq. (\ref{eq28}) is 
 still convergent, since given any $r_4 < s < \infty$ (no problems for $r\rightarrow\infty$), then 
 \begin{equation*}
     \int_{r_4}^{s}\dfrac{2Mra-a^2\lambda+acr^2}{\Delta \sqrt{R(r)}} dr \leq  \delta   M  F\big(x_s,k\big)  < \infty
 \end{equation*}
where $M$ is the maximum of the function $(2Mra-a^2\lambda+acr^2)/\Delta$ on the range of the integral and $x_s$ is the variable $x$ (see Appendix B) evaluated in $r=s$. Proceeding similarly to what was done for Eq. (\ref{eq-26}) and assuming $\varphi_a = 0$ and $q=0$ for both pulsar and observer (pulsar, black hole and observer lay on the equatorial plane, from Eq. (\ref{eq29}) we get
\begin{equation}\label{eq30}
\arraycolsep=1.8pt\def\arraystretch{2.6}
\begin{array}{l}
    -\varphi_e = \dfrac{a c }{\gamma}\delta \cdot \left[ F(\infty,k) \pm F(r_e,k) \right] \\
    + \dfrac{\delta a(a^2 c -2Mr_{+}+a\lambda\gamma)}{\gamma^2 (r_{-}-r_{+})(r_4-r_{+})} \Big[ \Big(1- \dfrac{l}{l_{+}} \Big) \Pi(\infty,l_{+},k) \\
     + \Big(\pm 1 \mp \dfrac{l}{l_{+}}\Big)\Pi (r_e,l_{+},k) + \dfrac{l}{l_{+}} F(\infty,k) \pm \dfrac{l}{l_{+}} F(r_e,k) \Big] \\
     + \dfrac{ \delta a \Big( 2Mr_{-}-a^2c-a\lambda \gamma \Big)}{ \gamma^2(r_{-}-r_{+})(r_4-r_{-}) } \Big[  \Big( 1 -\dfrac{l}{l_{-}} \Big)\Pi(\infty,l_{-},k) \\
     + \Big(\pm 1 \mp \dfrac{l}{l_{-}} \Big) \Pi(r_e,l_{-},k) + \dfrac{l}{l_{-}}F(\infty,k) \pm \dfrac{l}{l_{-}} F(r_e,k)   \Big] \\
     +\dfrac {2\lambda[ F(\infty ,k)\pm F(r_e,k)]}{\sqrt{(r_4-r_2)(r_3-r_1)}}
    \end{array}
\end{equation} 
where we choose  upper or lower sign for  flyby or direct trajectory, respectively. We also note that, in general, the angular integral in Eq. (\ref{eq29}) can be computed using 
\begin{equation*}
 \int_{0}^{u_{e,a}} \dfrac{\lambda}{\sin^2(\theta) \sqrt{\Theta(\theta)}} d\theta =  \pm  \dfrac{\lambda}{\sqrt{u_{-}}} \Pi\left(\sqrt{\dfrac{u_{e,a}}{u_{+}}},u_{+},\sqrt{\dfrac{u_{+}}{u_{-}}}  \right),
\end{equation*}\\
but, in the assumption of equatorial orbit (i.e. $\theta=\pi/2$) and using Eq. (\ref{eq28}), it simply reduces to a radial integral   
\begin{equation*}
  \int_{\gamma_{\theta}} \dfrac{\lambda}{\sin^2(\theta) \sqrt{\Theta(\theta)}} d\theta =   \int_{\gamma_{r}} \dfrac{\lambda}{\sqrt{R(r)}_{q=0}} dr 
\end{equation*}
and this gives the last term in Eq. (\ref{eq30}). \\
Generally, the particular case in which everything is restricted to the equatorial plane is simpler and, at the same time, more interesting, as this case corresponds to the strongest relativistic effects. Before schematizing the procedure we have followed, let us first explain how to relate metric coordinates to the coordinates of a pulsar orbiting a black hole. To this end, we will adopt the geometrical set-up already introduced in \cite{Eva_2019}. Due to the big difference in  mass between pulsar and (supermassive) black hole, we may consider the pulsar as a test particle, i.e. the center of mass coincide with the BH center. Of course, the pulsar will not remain, in general, in a fixed plane due to the frame dragging, and that the same orbit would rotate because of the relativistic precession of the periapsis. Nevertheless, we will assume a keplerian orbit in first approximation, with (almost all) relativistic effects encoded in post-Newtonian orbital parameters, as usual in pulsar timing models.  However, here we are not interested in individual post-Keplerian effects, since Eq. (\ref{eq27}) is a full relativistic formula for time delay; what we need to do is just to compare  formula (\ref{eq27}) with the analogue one without the surrounding matter ($c=0$, i.e. Kerr), in order to highlight the effect of the latter.
This is a difference with \cite{Bilel}, where the relativistic effect of frame dragging has been studied and compared to weak field post Newtonian approximations.

 In this case we can express the coordinate $(x,y,z)$ as follows: $x= r_{e}\cos(\omega+\phi)$, $y=r_{e}\sin(\omega+\phi)$, $z=0$, where $\omega$ is the argument of the periastron. A rotation around the $x$-axis by the inclination angle $i$ suffices to transform to the $(X,Y,Z)$ system. For the case of an edge-on equatorial pulsar orbit, that we will discuss later in the paper, the desired angle between pulsar and observer is then given by the angle $\vartheta$ in spherical coordinates $X= r\cos\psi\sin\vartheta$, $Y=r\cos\psi\sin\vartheta$, $Z=r\cos\vartheta$. In the common plane of pulsar and observer, the angle $\phi_{e}$ is then determined by $\varphi_{e}=\vartheta$ with $\cos\vartheta= -\sin\textit{i}\sin(\omega+\phi)$ and therefore
\begin{equation}\label{eq:phi} 
\cos\varphi_{e} = -\sin\textit{i}\sin(\omega+\phi)\,.
\end{equation} 


\section{Results}

First of all, we assume an extreme binary system of a pulsar orbiting a supermassive black hole with a mass of $M=4 \times 10^6 M_{\odot}$ (solar masses), i.e. $G M_{\odot} / c^2 = 1476 M$. Therefore, the propagation time delay will be expressed in seconds and the corresponding  adimensional value can be recovered by dividing by a factor $G M/c^3 \approx 19.7 s$. Different black hole masses (say $M_2$) lead to time delays multiplied by a factor $M_2/M$. \\

First, we notice that by choosing $c\approx 0$, the effects of matter are negligible and we are able to reproduce the results of \cite{Bilel} from our Eq. (\ref{eq27}). Just to give an example, we show a single plot for this case in Fig. (\ref{fig:PD-dark-energy}), where we also show the cases of a rotating and a not-rotating  Kiselev black holes in presence of dark energy. It turns out that while  the spin $a$  induces a shift in the phase of the maximum delay  without increasing or decreasing the delay, the effect of matter surrounding  the black hole results in a noticeable alteration of the delay, which  increases as the strength parameter $c$ increases. Interestingly, when the pulsar  is positioned between the observer and the black hole ($\theta \approx 0$) we predict a  slightly lower pulse delay than in the no-matter case. Difference with Schwarzchild and Kerr cases are less evident (but still appreciable) in presence of dust (see Fig. (\ref{fig:PD-dust})) and practically imperceptible in the case of pure radiation (see Fig. (\ref{fig:PD-radiation})).   \\

We also may  isolate the only effect of matter  by making  the subtraction
\begin{equation}\label{eq32}
    \Delta t_{matter} = \Delta t_{ex}(c) - \Delta t_{ex}(c=0)
\end{equation}
 where $\Delta t_{ex}(c=0)$ is obtained by putting $c=0$ in Eq. (\ref{eq27}). Here a comment is necessary. If one identifies $\Delta t_{ex}(c=0)$ in Eq. (\ref{eq32}) with the exact time delay in Kerr metric, then, in order to compare results derived in different space-times (Kiselev vs Kerr),  we first need to identify a physical invariant.  The common idea is to fix the circumference of a circle, which is an invariant characteristic. In particular, in Kerr space-time, such a circumference is given by $ \mathcal{C}_{kerr} = 2 \pi  \sqrt{r^2_{kerr}+a^2+2Ma^2/r_{kerr}}$, while  our metric (\ref{eq1}) leads to the (equatorial) circumference
\begin{equation}
  \mathcal{C}_{kis} =   2 \pi  \sqrt{r^2_{kis}+a^2+a^2 \left( \dfrac{2M}{r_{kis}} +c r_{kis}^{-1-3w}\right)}.
\end{equation}
Then the equality is achieved with
\begin{equation}\label{eq34}
    r_{kerr} = 2 \sqrt{-Q} \cos \left( \dfrac{\theta}{3}\right)
\end{equation}
where $Q$ depends on the type of matter
\begin{equation*}
    Q_{de}= -\dfrac{r^3_{kis}+a^2(2M+cr_{kis})}{3r_{kis}} , \; \; \; \;  Q_{dust}= -\dfrac{r^3_{kis}+a^2(2M+c)}{3r_{kis}}  \end{equation*}
    \begin{equation*}
    Q_{rad}= -\dfrac{r^4_{kis}+a^2(2Mr_{kis}+c)}{3r^2_{kis}} , \; \;  \theta = \arccos\Big(-\dfrac{Ma^2}{\sqrt{-Q^3}}\Big).
\end{equation*}
However, to make things easier, one could decide to use only Eq. (\ref{eq27}) to calculate the matter delay (\ref{eq32}), for both the contributions $\Delta t_{ex}(c)$, $\Delta t_{ex}(c=0)$.  This is also justified  by the fact that Eq. (\ref{eq34}) gives very convergent $r_{kerr}$, $r_{kis}$ values as long as $c$ is far from 1, as we are assuming. \\


\subsection{Effect of the black hole environment on the propagation time delay}


\begin{figure}
    \centering
    \includegraphics[width=0.48\textwidth]{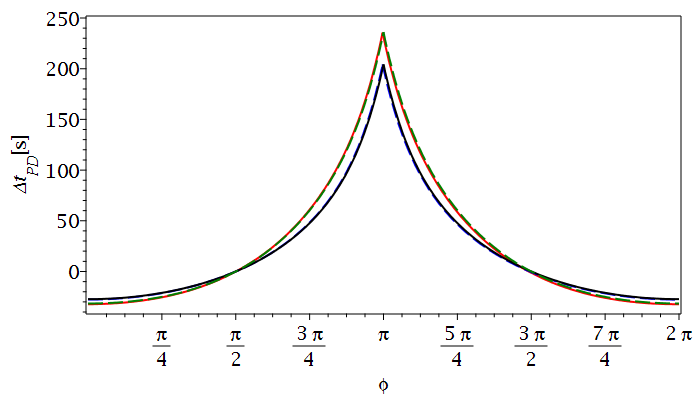}
    \includegraphics[width=0.48\textwidth]{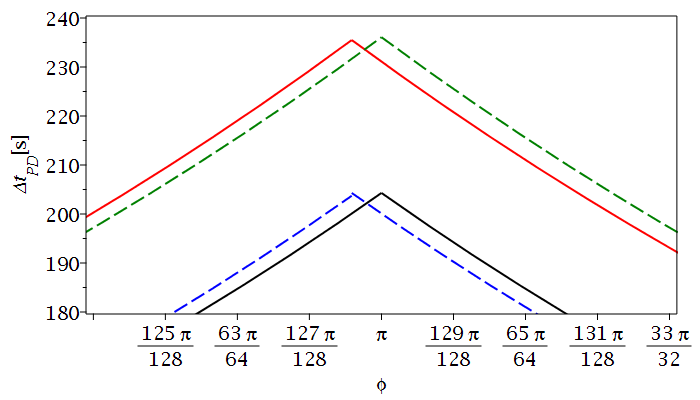}
  \includegraphics[width=0.48\textwidth]{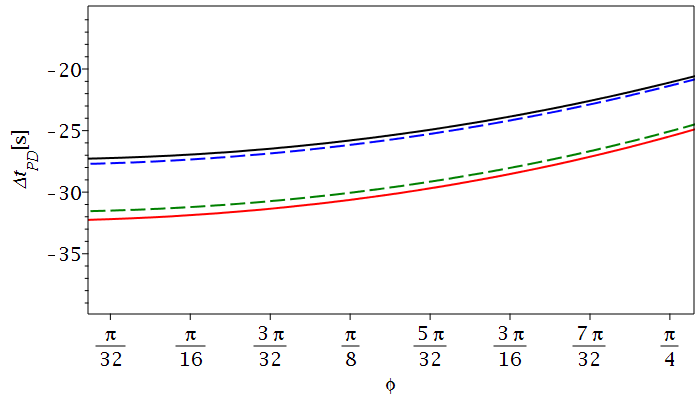}
    \caption{The exact time delay  $\Delta t_{\text{ex}}- \Delta t_{\text{R}}$ for a Schwazschild black hole (black line), a Kerr black hole with $a=0.9$ (dashed blue line), a non-rotating Kiselev black hole surrounded by dark energy with $c=0.01$ (dashed green line) and a rotating one with $a=0.9$ and $c=0.01$ (red line) for a circular edge-on orbit with a Schwarzschild radius $r_{\text{S}}=100\, M$. Bottom figures are a zoom of the top. At this distance, the additional time delay when the pulsar is behind the BH (superior conjunction) is more than 30 seconds, while it is reduced by a few seconds when the pulsar is in front of the BH (inferior conjunction). }
    \label{fig:PD-dark-energy}
\end{figure}

Given the Roemer delay $\Delta t_{\text{R}}$  which is given by  \cite{Blandford1976} :
\begin{equation}\label{eq:Roemer} 
\Delta t_{\text{R}} = \frac{A(1-\textit{e}^2)\sin\textit{i}\sin(\omega+\phi)}{c(1+\textit{e}\cos\phi)}
\end{equation} 
where $\textit{i}$ is the inclination of the orbital plane with respect to the plane of sky and $\omega$ is the argument of periapsis, $\phi$ is the argument of the pulsar's position and $\textit{e}$ is the eccentricity of the orbit, we show in figure \ref{fig:PD-dark-energy}, the time delay  $\Delta t_{\text{ex}}- \Delta t_{\text{R}}$ (see equation \ref{eq27})  for a Schwarzschild and  a Kerr black hole both in the case where they are in a vacuum as well as surrounded by dark energy with $c=0.01$. We choose a simple pulsar trajectory of an edge-on ($\mathit{i} = \pi/2$) circular orbit. The ascending node with respect to the plane of sky is used as the referenece point i.e $\varphi_{\text{ref}} = \pi/2$, which with $\omega = -\pi/2$ simplifies to   $\varphi_{\text{ref}} = \phi_{\text{ref}} = \pi/2$. Note that all exact propagation time delays  $\Delta t_{\text{ex}}$  includes considerable offset. Therefore we correct them by adding global constants to the individual delays such that they exactly vanish at $\phi_{\text{ref}} = \pi/2$. \\
As expected for a circular edge-on orbit in spherical symmetry, the curve representing the Schwarzschild black hole ($a=0,c=0$) is symmetric with respect to the superior conjunction at $\phi=\pi$. Once the black hole rotates this symmetry is broken with a slight shift in the top of the curve which corresponds to the switch from contra-rotating to co-rotating lightlike geodesics emitted from the pulsar to the observer which was already discussed in details in \cite{Bilel}. However, if the dark energy is present ($w=-1/3$), a considerably additional amount of  time delay (more than 30 seconds if the pulsar is at a distance $r_s=100$ M) is induced if the pulsar comes behind the black hole along its trajectory. On the other side, dark energy contributes negatively to the time delay (with an advance of a few seconds) once the pulsar is at the front of the observer. This behaviour is seen both for a Schwarzschild and a Kerr black hole as it is illustrated in the bottom figure \ref{fig:PD-dark-energy}.

\begin{figure}
    \centering
    \includegraphics[width=0.48\textwidth]{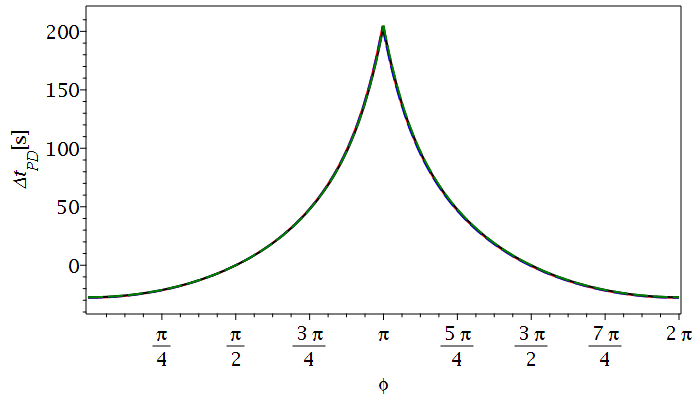}
    \includegraphics[width=0.48\textwidth]{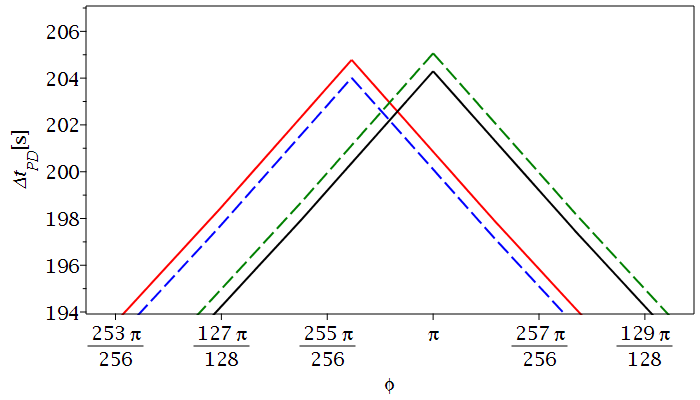}
    \includegraphics[width=0.48\textwidth]{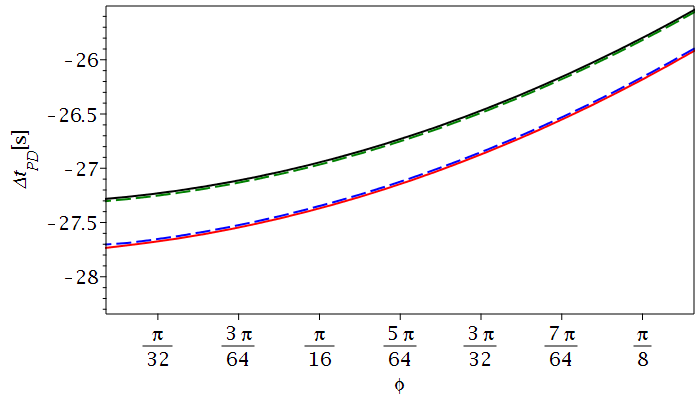}
    \caption{The exact time delay  $\Delta t_{\text{ex}}- \Delta t_{\text{R}}$ for a Schwarzschild black hole (black line), a Kerr black hole with $a=0.9$ (dashed blue line), a non-rotating Kiselev black hole surrounded by dust with $c=0.01$ (dashed green line) and a rotating one with $a=0.9$ and $c=0.01$ (red line) for a circular edge-on orbit with a Schwarzschild radius $r_{\text{S}}=100\, M$. Bottom figures are a zoom of the top.}
    \label{fig:PD-dust}
\end{figure}

Surprisingly, in the case the dust is present at the vicinity of the black hole ($w=0$), the same above feature is observed as in the case of dark energy but with a very small amplitude (see figure \ref{fig:PD-dust}). This finding appears to contradict the expected behaviour, in which the dust would induce a constant time delay for a circular pulsar orbit, regardless of its position with respect to the black hole. 

\begin{figure}
    \centering
    \includegraphics[width=0.48\textwidth]{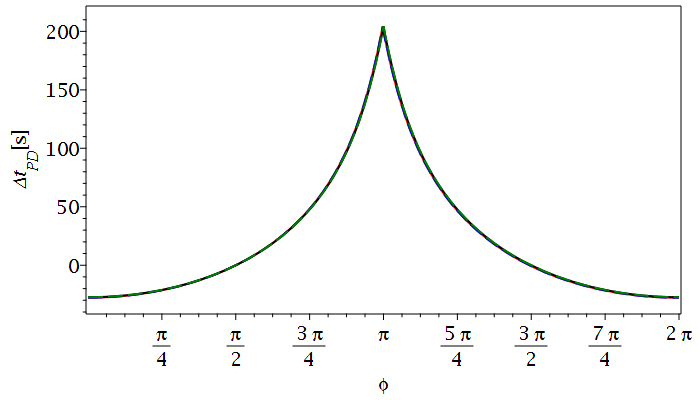}
    \includegraphics[width=0.48\textwidth]{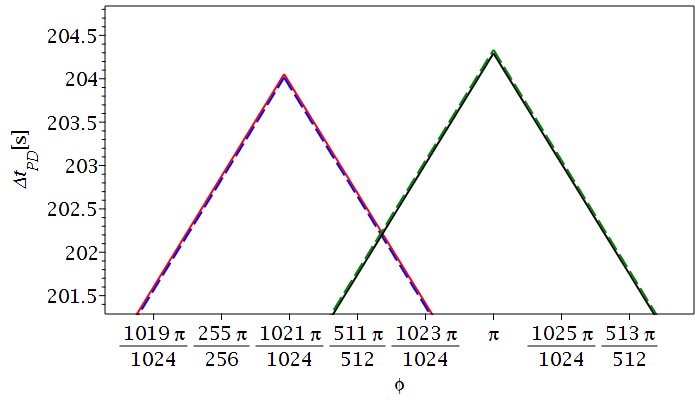}
        \includegraphics[width=0.48\textwidth]{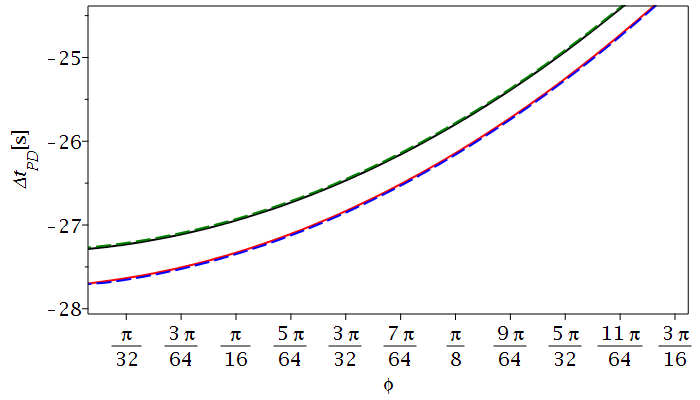}
    \caption{The exact time delay  $\Delta t_{\text{ex}}- \Delta t_{\text{R}}$ for a Schwazschild black hole (black line), a Kerr black hole with $a=0.9$ (dashed blue line), a non-rotating Kiselev black hole surrounded by radiation with $c=0.01$ (dashed green line) and a rotating one with $a=0.9$ and $c=0.01$ (red line) for a circular edge-on orbit with a Schwarzschild radius $r_{\text{S}}=100\, M$. Bottom figures are a zoom of the top.}
    \label{fig:PD-radiation}
\end{figure}

In figure \ref{fig:PD-radiation}, the propagation time delay of light geodesics emitted from the pulsar in  the presence of radiation ($w= 1/3$) is investigated. As it is expected, the radiation contributed positively to the propagation time delay along the pulsar trajectory around the black hole. However this contribution is maximal once the pulsar is at a superior conjecture and minimal at the inferior conjecture. 


\section{Conclusions}

Most applications of pulsars involve a technique called “pulsar timing”, i.e. the measurement of the time of arrival (ToA)
of photons emitted by the pulsar, which are then compared with a theoretical model. With a collection of ToAs in hand, it becomes possible to fit a model of the pulsar’s timing behaviour, accounting for every rotation of the neutron star. Depending on whether the pulsar is isolated or binary, the multi-parameter fit gives  several important parameters (so-called ephemeris), like period, period derivative, orbital period (if binary), position in the sky, eccentricity, ecc. Over the years there have been numerous studies on possible violations of General Relativity (GR) using the timing of the pulsars  \citep[e.g.][]{Stairs_2003,https://doi.org/10.48550/arxiv.2204.13468}, revealing that pulsars are a great research tool in this field, offering also the possibility  to test the no-hair theorem as well as the
cosmic censorship conjecture \citep{Liu_2012,Izmailov:2019cqr}. 
More recently, pulsar timing is also used to detect and characterize the low-frequency gravitational wave universe through timing an array of approximately 100 millisecond pulsars (MSP) using the largest radio telescopes in the world. Usually, the timing model to predict ToAs is based on Damour and Deruelle’s
approach using a post-Newtonian expansion to treat the relativistic
two body problem \citep{deruelle}, and its corresponding
relativistic effects are described by a set of post-Keplerian parameters. However, the validity of the post-Newtonian approximation (that assumes a weak
field) it is no longer guaranteed for a pulsar orbiting closely a SMBH, in particular if  
 pulsar, black hole and observer are (nearly) aligned. This means that full general relativistic computations are sometimes necessary.  \\
In this paper we study, from an analytical point of view the further effect of the presence of matter (radiation, dust, dark energy) in the surrounding of the central massive black hole. Hence, the aim of this work was to investigate if and  how the  propagation time delay is affected by different types of matter, whose presence is certainly not negligible in the case of SgrA$^{*}$. We have modeled such a presence with the (rotational) Kiselev metric, which in addition to the spin parameter $a$, depends on the strength parameter $c$, as well as on the state parameter $w$ depending on the type of matter ($w=\pm 1/3, 0$). Recently, evidence has emerged that black holes could contribute to the dark energy content of the universe, resulting cosmologically coupled \citep{BH_DE_inside,BH_DE_inside_2}; therefore considering the possibility of dark energy around SMBHs could give clues in this direction. \\
 We first computed geodesic equations in a general, rotational Kiselev metric and characterized the corresponding parametr space $(\lambda,q$), where $\lambda$ is the (adimensional) angular momentum and $q$ is the (adimensional) Carter constant, finding that generally the presence of matter enlarge the region of direct trajectories  of photons moving from the pulsar and a far observer (see Fig. \ref{fig:1}), at the expense of a flyby. This effect is more evident for "dark energy" and less evident for radiation, and the divergence with Kerr case ($c=0$) increases as the value of $c$ increases, as we expected.  Then, we analytically solve motion equations using the Mino time and elliptic integrals, following the strategy used in \cite{Bilel}. Our main formula, Eq. (\ref{eq-26}), reproduce Kerr case when $c=0$ (regardless of the value of $w$). To estimate the magnitude of the effect, we assumed a Keplerian orbit for the pulsar orbiting a supermassive black hole in the equatorial plane ($q=0$), where relativistic effect are strongest. Due to the big difference mass between pulsar and BH, we considered the pulsar as a test particle. We did not deal with individuals post-Keplerian  effects, but just compared our model to the Kerr case in order to isolate the effect of matter. We found that a deviation is real and very pronounced in presence of dark energy ($w=-1/3$):  while  the spin $a$ only induces a shift in the phase of the maximum delay  without increasing or decreasing the delay, the effect of matter surrounding  the black hole results in a noticeable alteration of the delay, which  increases as the strength parameter $c$ increases. For example, if the pulsar is at a distance $r_s= 100$ M, the additional time
delay when the pulsar is behind the BH (superior conjunction) would be more than 30 seconds. Interestingly, when the pulsar  is positioned between the observer and the black hole ($\phi \approx 0$, i.e. inferior conjunction), we predict a  slightly lower pulse delay (with an advance of a few seconds)  than in the no-matter case (see Fig. (\ref{fig:PD-dark-energy})). Differences with Schwarzchild and Kerr cases are less evident (but still appreciable) in presence of dust (see Fig. (\ref{fig:PD-dust})) and practically imperceptible in the case of pure radiation (see Fig. (\ref{fig:PD-radiation})).

\begin{figure}
    \centering
    \includegraphics[width=0.48\textwidth]{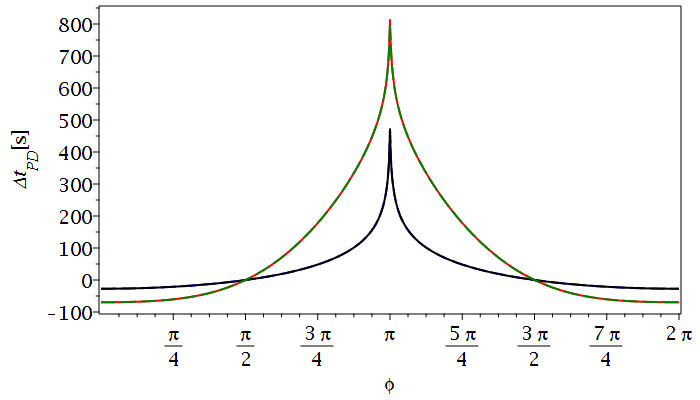}
    \includegraphics[width=0.48\textwidth]{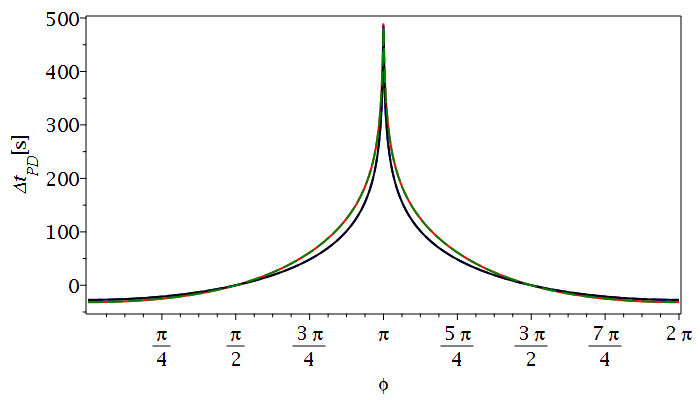}
    \caption{(Top) The exact time delay ($\Delta t_{\text{ex}}- \Delta t_{\text{R}}$) for SGR J1745-2900,  the closest magnetar orbiting SgrA*,  for a Schwarzschild black hole (black line), a Kerr black hole with $a=0.9$ (dashed blue line), a non-rotating Kiselev black hole surrounded by dark energy with $c=10^{-4}$ (dashed green line) and a rotating one with $a=0.9$ and $c=10^{-4}$ (red line) for a circular edge-on orbit with a Schwarzschild radius $r_{\text{S}}=10^5 \, M$, where $M=4 \times 10^6 $ $M_{\odot}$. (Bottom) As before, but with $c=10^{-5}$. By comparing the observed magnetar delays at superior and inferior conjunctions with theoretical predictions, a signature of the presence of  dark energy could appear.  }
    \label{fig:finale}
\end{figure}

Even if at the moment no pulsars close enough to SgrA* are known, their search is attracting more and more efforts from the scientific community \cite{new_pulsars}, due to the enormous implications that such a discovery would have \cite{probe_2004}. Since our delay formulas do not depend on the energy of the emitted photons, they are also valid for magnetars, which often show variability at higher 
energies than radio waves. This allows us to estimate the delay of SGR J1745-2900,  the closest magnetar orbiting SgrA*. At a distance $r_s = 0.1 $ pc \cite{distance_2015}, i.e. $r_s \simeq 5 \times 10^5 M $, and assuming an edge-on orbit \footnote{More observations are needed to determine the orientation of SGR J1745-2900's orbit.}, it turns out that the  difference in delays between the no-matter case  (Kerr metric  with $a=0.9$) and the dark-energy case (Kiselev metric with $a=0.9$, $w=-1/3$ and $c=10^{-4}$) is approximately  \textit{larger} than $300$ seconds at superior conjunction and \textit{shorter} than $40$ seconds  at inferior conjunction. When $c=10^{-5}$ differences flatten significantly and are almost imperceptible (see Fig. \ref{fig:finale}). For completeness, when $c=10^{-3}$ the delay in presence of dark energy   would be  $\sim 4000$ seconds greater, a value so large as to make us consider $c=10^{-3}$ unlikely at that distance from SgrA*. Finally, we notice that, at the considered distance, the spinning effects are almost completely absent almost completely absent and this makes the search for matter effects even cleaner.   By comparing the observed magnetar delays at superior and inferior conjunctions with theoretical predictions, a signature of the presence of matter (dark energy in particular) could appear. However,  at the moment the timing of SGR J1745-2900  has not yet reached sufficient levels of precision for this purpose, mainly due to the intrinsic variability of the source.

Beyond the information on any matter present and its effects on timing, the advantage of this work, as compared to numrical ray-tracing methods, is the possibility to isolate diverging terms providing at the same time non-approximate results. Besides, our formulas could be integrated into a new relativistic timing model that is suitable for extreme binary systems where the presence of matter is not-negligible. However, this could be a not-easy task, since it requires to know the environment around the black hole. Conversely, one could exploit pulsar timing to constrain $w$ in the vicinity of the black hole, if the pulsar is close enough. The difficulty in this case would be isolating the matter effect from all other effects, which is impossible in strong field situations, where the (non-linear) full relativistic equations must be used. \\
 Possible extensions of this work include the study of non-equatorial orbits, the addition of a pulsar's spin, the treatment of the pulsar as a timelike geodesic.  Parallel works, on the other hand, may concern the use of other metrics, such as the Kerr-Newmann and  Kerr-Sen  ones. In principle, the equations of motion for lighlike geodesics are solvable in terms of elliptic or hyperelliptic  integrals. Such investigations would probe the possibility to test the no-hair theorem by predicting the time delays induced by the additional parameters (axion, charge, ecc...). A degeneracy between different parameters, however,  could  arise, invalidating an accurate measurement of single parameters. In particular, in light of the results found here, the presence of matter surrounding the black hole could affect accurate measurements of the spin parameter $a$ by pulsar timing methods.  \\
 In conclusion, our results could serve, in principle, to better understand the environment in which the binary system is located by exploiting the timing of the pulsar, provided that this effect is separable from the others.

\begin{acknowledgments}
A.C. would acknowledge the support by the  Istituto Nazionale di Fisica Nucleare (INFN) {\it Iniziativa Specifica} QGSKY. Furthermore, A.C. would express his gratitude to  Prof. Alberto Sesana for  ospitality at University of Milano 'Bicocca' and Prof. Andrea Possenti for hospitality at INAF-Cagliari, as well as for their valuable advice. 
\end{acknowledgments}







\appendix

\section{Multiple Roots}
In this section we report quadruple, triple and double roots for the radial potential $R(r)$ for a Kiselev metric, in presence of dust and radiation. We will omit the computation of simple roots, since it is entirely equivalent to Eq. (\ref{eq15}). Furthermore, quadruple roots  remain unchanged and, hence, equal to the Kerr metric.  \\
For radiation ($w=1/3$), triple roots are
\begin{equation*}
    q= \dfrac{8r^3}{2M}-(a+\lambda)^2 , \; \; \; \; \; \lambda= a+ \dfrac{r^2}{a}\left( 3-\dfrac{2r}{M} \right)
\end{equation*}
with the only non-zero real radial coordinate
\begin{equation*}
   r = M- \left[ M \left( M^2-a^2+c \right) \right]^{1/3}.
\end{equation*}
Double solutions are obtained for
\begin{equation}
    q= \dfrac{r^2}{a^2(M-r)^2}\left[ 4a^2 \left(Mr+c \right)- \left( 3Mr-r^2+2c \right)^2 \right] ,
\end{equation}
\begin{equation*}
    \lambda =   a+ \dfrac{r^2}{a}\left( 1 - \dfrac{2 \Delta}{r(r-M)} \right).
\end{equation*}
On the boundary of the allowable region (\ref{eq14}), $q=0$ or $q=-(\lambda \pm a)^2$. The first condition is realized for four real and distinct radial values, two of which are always positive. The second condition ($q=-(\lambda + a)^2$) give 
\begin{equation*}
 r= \dfrac{3M^2+c}{3M}\cos \left[ \dfrac{1}{3} \theta_{1} +\dfrac{2\pi k}{3}\right] - \dfrac{A}{3}
\end{equation*}
where $\theta_1$ and $A$ satisfy 
\begin{equation*}
    1 + \cos\theta_{1} = \dfrac{54M^4 (M^2-a^2+c)}{3M^2+c}^3  , \; \; \; \; \; A=\dfrac{c-3M^2}{2M}.
\end{equation*}
As in the Kerr case, the last condition ($q=-(\lambda - a)^2$) implies that $r=r_{\pm}$, where $r_{\pm}$ is the radial position of inner and outer horizons. \\
In presence of dust ($w=0$),  for triple roots we get
\begin{equation*}
    q=\dfrac{r^3}{a^2}\left[ \dfrac{8a^2(c+2M)-r(6M+3c-4r)^2}{(2M+c)^2} \right] ,
\end{equation*}
\begin{equation}
    \lambda = a + \dfrac{r^2}{a}\left( \dfrac{6M+3c-4r}{2M+c} \right)
\end{equation}
as well as
\begin{equation*}
r= \dfrac{1}{2} \left[ 2M+c + n^{1/3} \right].
\end{equation*}
with $n=\left(4a^2c-c^3+8a^2M-6c^2M-12cM^2-8M^3 \right)$. Finally, double solutions  occur when 
\begin{equation*}
    q= \dfrac{a^2 \left[ 16a^2\Delta - \left( 4\Delta + r(2M-2r+c) \right)^2 \right]}{a^2\left( 2M-2r+c \right)^2} ,
\end{equation*}
\begin{equation*}
    \lambda = a + \dfrac{r^2}{a} - \dfrac{4r \Delta}{a (2r-2M-c)}
 \end{equation*}
which leads to two cubic equations when $q=0$ and $q=-(\lambda+a)^2$, as well as $r=r_{\pm}$ when $q=-(\lambda-a)^2$. Their solutions are, respectively  
\begin{equation*}
    r= 2 \sqrt{M(c+M)}\cos \left( \dfrac{\theta_{2}}{3} + \dfrac{2 \pi k}{3} \right) + 2M+c ,
\end{equation*}
\begin{equation}
 r = \dfrac{c+2M}{2}\cos \left( \dfrac{\theta_{3}}{3} + \dfrac{2 \pi k}{3} \right)   + \dfrac{2M+c}{4} 
\end{equation}
where
\begin{equation*}
\begin{aligned}
    \cos\theta_{2} &= \dfrac{(c+2M)(2a^2+2c^2-cM-M^2)}{\left( M^2+cM+c^2 \right)^{3/2}} \;, \\
    \cos\theta_{3} &=1-\dfrac{8a^2}{(c+2M)^2}.
    \end{aligned}
\end{equation*}

\section{Time delay integrals}
In the following, we summarize the computation of the time delay integral (\ref{eq-26}) in presence of a dark energy component. We also report the final formulas for dust and radiation, the steps being completely analogous. For the angular integral (\ref{eq25}) we refer to \cite{Bilel}. \\
\subsection{Dark Energy}

To solve analytically the integral (\ref{eq22}), we start with the cahnge of variable:
\begin{equation}
    x^2 = \dfrac{(r-r_4)(r_3-r_1)}{(r-r_3)(r_4-r_1)}
\end{equation}
where $r_i$ are the roots of the radial potential $R(r)$. This leads the integral into the form 
\begin{equation}
\arraycolsep=2.8pt\def\arraystretch{2.6}
\begin{array}{l}
    T(r,\lambda,q;c) = \displaystyle\int_{r_4}^{r} \dfrac{G(r)}{\Delta \sqrt{R(r)}} dr = \\
    \delta \cdot\displaystyle\int_{0}^{x(r)}\dfrac{G(x)}{\Delta(x)\sqrt{(1-x^2)(1-k^2x^2)}} dx
    \end{array}
\end{equation}
where $G(r)$ is defined in Eq. (\ref{eq21}) and 
\begin{equation*}
    k^2 = \dfrac{(r_3-r_2)(r_4-r_1)}{(r_3-r_1)(r_4-r_2)}.
\end{equation*}
We then expand the quantity $\dfrac{G(x)}{\Delta}$ in partial fractions:
\begin{equation*}
\dfrac{G(x)}{\Delta} =    \dfrac{1}{\gamma^3} \Big[ 4M^2+(2Mr-a^2c^2)\gamma+(r^2-ac\lambda)\gamma^2 \Big] +  A(r)+B(r)
\end{equation*}
where we have defined 

\begin{equation*}
A(r)=\frac{8 M^3r_{+}-4 a^2 M \left(M+c r_{+}\right)+a^3 c \lambda-2 \Tilde{a} r_{+} M \lambda}{2 \sqrt{M^2-a^2\gamma} \left(r-r_{+}\right) \gamma^3} ,
\end{equation*}

\begin{equation*}
 B(r)=\frac{-8 M^3 r_{-}+4 a^2 M\left(M+c r_{-}\right)-a^3 c \lambda+2 \Tilde{a} M \lambda r_{-}}{2 \sqrt{M^2-a^2\gamma} \left(r-r_{-}\right)  \gamma^3}   .
\end{equation*}
where $\Tilde{a}=a(1-2c)$. We then use the following well-known closed integrals 
\begin{equation*}
    F(x,k) = \displaystyle\int_{0}^{x} \dfrac{dx}{\sqrt{(1-x^2)(1-k^2x^2)}} ,
\end{equation*}
\begin{equation*}
    E(x,k) = \displaystyle\int_{0}^{x} \dfrac{\sqrt{1-k^2x^2}}{\sqrt{1-x^2}} dx  ,
\end{equation*}
 \begin{equation*}
    \Pi(x,l,k) = \displaystyle\int_{0}^{x} \dfrac{dx}{(1-lx^2)\sqrt{(1-x^2)(1-k^2x^2)}} ,
\end{equation*}  
 having defined the constant $l= \dfrac{r_1-r_4}{r_1-r_3}$. Noting that  
 \begin{equation*}
     \displaystyle\int_{0}^{x} \dfrac{-x^2}{(1-lx^2)\sqrt{(1-x^2)(1-k^2x^2)}} = \dfrac{1}{l} \Big[ F(x,k) - \Pi(x,l,k)\Big] ,
 \end{equation*}
 and since \citep{higher}
\begin{equation*}
\begin{aligned}
    &\dfrac{1}{(1-lx^2)^2 y(x)} = \\
    &C_1 \dfrac{d}{dx} \Big[ \dfrac{x  y(x)}{1-lx^2} \Big] + \dfrac{C_2}{y(x)} + \dfrac{C_3 (1-k^2x^2)}{y(x)} + \dfrac{C_4}{y(x) (1-lx^2)} 
\end{aligned}
\end{equation*}
(to be integrated), one obtains Eq. (\ref{eq-26}) (for $w=-1/3$), where we also called 
\begin{equation}
B_{+}=\frac{8 M^3r_{+}-4 a^2 M \left(M+c r_{+}\right)+a^3 c \lambda-2 \Tilde{a} r_{+} M \lambda}{2 \sqrt{M^2-a^2\gamma} \left(r_4-r_{+}\right) \gamma^3} ,
\end{equation}
\begin{equation}
 B_{-}=\frac{-8 M^3 r_{-}+4 a^2 M\left(M+c r_{-}\right)-a^3 c \lambda+2 \Tilde{a} M \lambda r_{-}}{2 \sqrt{M^2-a^2\gamma} \left(r_4-r_{-}\right)  \gamma^3} 
\end{equation}
where 
\begin{equation*}
    r_{\pm} = \dfrac{M\pm \sqrt{M^2-a^2\gamma}}{\gamma}
\end{equation*}
are the positions of the horizons in the case $w=-1/3$.
The presence of diverging parts for $r\rightarrow \infty$ in $\Pi(x,l,k)$ and $\dfrac{\sqrt{R(r)}}{r-r_3}$ can be isolated  by exploiting the relation
\begin{equation}
    \Pi(x,l,k) = F(x,k)-\Pi\Big(x,\frac{k^2}{l},k\Big) + \dfrac{\ln(Z)}{2P}
\end{equation}
with
\begin{equation}
\begin{aligned}
Z & =\frac{\sqrt{\left(1-x^2\right)\left(1-k^2 x^2\right)}+P x}{\sqrt{\left(1-x^2\right)\left(1-k^2 x^2\right)}-P x}, \\
P^2 & =\frac{(l-1)(l-k^2)}{l}=\frac{\left(r_3-r_4\right)^2}{\left(r_4-r_2\right)\left(r_3-r_1\right)} .
\end{aligned}
\end{equation}
Therefore, Eq. (\ref{eq-26}) becomes
\begin{equation}
\arraycolsep=1.8pt\def\arraystretch{2.6}
\begin{array}{l}
T_{r}\left(r, \lambda,q;c\right)= \\
\delta \cdot \Big[F(x, k) \cdot \Big(4 M^2 \gamma^{-3}-a^2 c^2 \gamma^{-2}-a c \lambda \gamma^{-1}+2 M r_3 \gamma^{-2} \\
+\frac{1}{2} \gamma^{-1}\big[r_1\big(r_3-r_4\big)+r_3\big(r_3+r_4\big)\big]+\frac{B_{+} l}{l_{+}}+\frac{B_{-}  l}{l_{-}}
+ 2 M (r_4-r_3) \gamma^{-2} \Big) \\
+E\big(x,k\big) \Big(-\frac{1}{2} \gamma^{-1} \big(r_4-r_2\big)\big(r_3-r_1\big)\Big)-\\
-\Pi\big(x, \frac{k^2}{l}, k\big) \Big(2 M r_4 \gamma^{-2}-2 M r_3 \gamma^{-2}\Big)+ \\
+\Pi\big(x,l_{+}, k\big) \Big(B_{+}-\frac{l B_{+}}{l_{+}}\Big)+\Pi\big(x, l_{-}, k\big) \Big(B_{-}-\frac{l B_{-}}{l_{-}}\Big)\Big]+ \\
+2M\delta\gamma^{-2} \big(r_4 - r_3 \big) \dfrac{\ln (Z)}{2 P} 
+\dfrac{\sqrt{R}}{r-r_3} \gamma^{-1}. \\
\end{array}
\end{equation}
where the last two terms are divergent when $r\rightarrow\infty$:
\begin{equation}
    \dfrac{\sqrt{R(r)}}{r-r_3} = r + r_3 + \mathcal{O}\left(\dfrac{1}{r}\right)
\end{equation}
\begin{equation}
    \ln(Z) = \ln \left( \dfrac{2}{r_3+r_4} \right)+ \ln(r) + \mathcal{O}\left( \dfrac{1}{r} \right)
\end{equation}
For the angular integral (\ref{eq25}) as well as for the constants $v$ and $w$ we directly refer to Eqs. (A17-A19) in \cite{Bilel}. \\

\subsection{Radiation}
In a very similar way we obtain for the radiation case ($w=1/3$)
\begin{equation}
\arraycolsep=1.8pt\def\arraystretch{2.6}
\begin{array}{l}
T_{r}\left(r, \lambda,q;c\right)=\delta \cdot \Big[F(x,k) \Big( 4M^2+c+2Mr_3 \\
+ \dfrac{1}{2}\big[r_1(r_3-r_4) 
+r_3(r_3+r_4) \big]  +2M (r_4-r_3) \\ + \dfrac{B_{+}l}{l_{+}} +\dfrac{B_{-}l}{l_{-}}\Big) 
+E(x,k)\Big( -\dfrac{1}{2}(r_4-r_2)(r_3-r_1) \Big) \\
 -\Pi\Big(x,\dfrac{k^2}{l},k\Big) \Big( 2M (r_4-r_3)\Big)  \\
+\Pi(x,l_{+},k)\Big( B_{+} -\dfrac{B_{+}l}{l_{+}} \Big) + \Pi(x,l_{-},k)\Big( B_{-} -\dfrac{B_{-}l}{l_{-}} \Big) \Big]\\
+ 2M\delta  (r_4-r_3)\dfrac{\ln{Z}}{2P} +\dfrac{\sqrt{R(r)}}{r-r_3}
\end{array}
\end{equation}
where 
\begin{equation}
\begin{aligned}
    R(r)= \Big( r^2-2M&r+a^2- c\Big)\Big( -q-\lambda^2-a^2+2a\lambda\Big)  \\
    &+\Big(r^2+a^2-\lambda a \Big)^2 \; ,
    \end{aligned}
\end{equation}
\begin{equation*}
    r_{\pm} = M \pm \sqrt{M^2-a^2+c} \; ,
\end{equation*}
\begin{equation*}
    B(r) = \dfrac{c^2+4M^2 (c-a^2)+4cMr+8M^3r-a\lambda(c+2 M r)}{ 2(r_4-r)\sqrt{M^2-a^2+c}  } ,
\end{equation*}
\begin{equation}
    B_{+} = B(r_{+}) , \; \; \; \; \; B_{-} = - B(r_{-}).
\end{equation}
Finally, the $\varphi$ coordinate is given in the integral form
\begin{equation}
    \varphi_{a}-\varphi_{e}  = \int_{\gamma_r}\dfrac{2Mra-a^2\lambda+ac}{\Delta\sqrt{R(r)}} dr + \dfrac{2 \lambda F(x,k)}{\sqrt{(r_4-r_2)(r_3-r_1)}}.
\end{equation}

\subsection{Dust}
As done in the previous section, the time delay in presence of dust ($w=0$) reads as 
\begin{equation}
\arraycolsep=1.8pt\def\arraystretch{2.6}
\begin{array}{l}
T_{r}\left(r, \lambda,q;c\right)=\delta \cdot \Big[F(x,k) \Big( 4M^2+4cM+c^2+r_3(2M+c) \\
+(2M+c)(r_4-r_3) 
+\dfrac{1}{2}\big[ r_1(r_3-r_4)+r_3(r_3+r_4)\big] + \dfrac{B_{+}l}{l_{+}} + \dfrac{B_{-}l}{l_{-}}\Big) \\
+E(x,k)\Big(-\dfrac{1}{2}(r_4-r_2)(r_3-r_1) \Big) \\
-\Pi\big(x,\dfrac{k^2}{l},k\big)(2M+c)(r_4-r_3) \\
+\Pi(x,l_{+},k)\Big( B_{+}-\dfrac{lB_{+}}{l_{+}}\Big) + \Pi(x,l_{-},k)\Big( B_{-}-\dfrac{lB_{-}}{l_{-}}\Big)\Big]\\
+\delta (2M+c)(r_4-r_3) \dfrac{\ln{Z}}{2P} +\dfrac{\sqrt{R(r)}}{r-r_3}
\end{array}
\end{equation}

where now 
\begin{equation}
    R(r)= -\Big( r^2-2Mr+a^2-c r\Big)\Big( q+\lambda^2+a^2-2a\lambda\Big)+\Big(r^2+a^2-\lambda a \Big)^2 ,
\end{equation}
\begin{equation*}
    r_{\pm} =\dfrac{1}{2}\Big[ c+2 M \pm \sqrt{4M^2-4a^2+4cM+c^2}\Big],
\end{equation*}
\begin{equation*}
    B(r) = \dfrac{(c+2M)\Big[-a^2(c+2M)+(c+2M)^2r-a\lambda r \Big]}{ (r_4-r) \sqrt{(2M+c)^2-4a^2}  } ,
\end{equation*}
\begin{equation}
    B_{+} = B(r_{+}) , \; \; \; \; \; B_{-} = - B(r_{-}).
\end{equation}
Finally, the $\varphi$ coordinate is given in the integral form
\begin{equation}
    \varphi_{a}-\varphi_{e}  = \int_{\gamma_r}\dfrac{(2Ma+ac)r-a^2\lambda}{\Delta\sqrt{R(r)}} dr + \dfrac{2 \lambda F(x,k)}{\sqrt{(r_4-r_2)(r_3-r_1)}}.
\end{equation}


\begin{thebibliography}{32}%
\makeatletter
\providecommand \@ifxundefined [1]{%
 \@ifx{#1\undefined}
}%
\providecommand \@ifnum [1]{%
 \ifnum #1\expandafter \@firstoftwo
 \else \expandafter \@secondoftwo
 \fi
}%
\providecommand \@ifx [1]{%
 \ifx #1\expandafter \@firstoftwo
 \else \expandafter \@secondoftwo
 \fi
}%
\providecommand \natexlab [1]{#1}%
\providecommand \enquote  [1]{``#1''}%
\providecommand \bibnamefont  [1]{#1}%
\providecommand \bibfnamefont [1]{#1}%
\providecommand \citenamefont [1]{#1}%
\providecommand \href@noop [0]{\@secondoftwo}%
\providecommand \href [0]{\begingroup \@sanitize@url \@href}%
\providecommand \@href[1]{\@@startlink{#1}\@@href}%
\providecommand \@@href[1]{\endgroup#1\@@endlink}%
\providecommand \@sanitize@url [0]{\catcode `\\12\catcode `\$12\catcode
  `\&12\catcode `\#12\catcode `\^12\catcode `\_12\catcode `\%12\relax}%
\providecommand \@@startlink[1]{}%
\providecommand \@@endlink[0]{}%
\providecommand \url  [0]{\begingroup\@sanitize@url \@url }%
\providecommand \@url [1]{\endgroup\@href {#1}{\urlprefix }}%
\providecommand \urlprefix  [0]{URL }%
\providecommand \Eprint [0]{\href }%
\providecommand \doibase [0]{https://doi.org/}%
\providecommand \selectlanguage [0]{\@gobble}%
\providecommand \bibinfo  [0]{\@secondoftwo}%
\providecommand \bibfield  [0]{\@secondoftwo}%
\providecommand \translation [1]{[#1]}%
\providecommand \BibitemOpen [0]{}%
\providecommand \bibitemStop [0]{}%
\providecommand \bibitemNoStop [0]{.\EOS\space}%
\providecommand \EOS [0]{\spacefactor3000\relax}%
\providecommand \BibitemShut  [1]{\csname bibitem#1\endcsname}%
\let\auto@bib@innerbib\@empty
\bibitem [{\citenamefont {Taylor}\ and\ \citenamefont
  {Weisberg}(1989)}]{taylor1989New}%
  \BibitemOpen
  \bibfield  {author} {\bibinfo {author} {\bibfnamefont {J.~H.}\ \bibnamefont
  {Taylor}}\ and\ \bibinfo {author} {\bibfnamefont {J.~M.}\ \bibnamefont
  {Weisberg}},\ }\bibfield  {title} {\bibinfo {title} {Further experimental
  tests of relativistic gravity using the binary pulsar psr 1913+ 16},\
  }\href@noop {} {\bibfield  {journal} {\bibinfo  {journal} {The Astrophysical
  Journal}\ }\textbf {\bibinfo {volume} {345}},\ \bibinfo {pages} {434}
  (\bibinfo {year} {1989})}\BibitemShut {NoStop}%
\bibitem [{\citenamefont {Stairs}(2003)}]{Stairs_2003}%
  \BibitemOpen
  \bibfield  {author} {\bibinfo {author} {\bibfnamefont {I.~H.}\ \bibnamefont
  {Stairs}},\ }\bibfield  {title} {\bibinfo {title} {Testing general relativity
  with pulsar timing},\ }\bibfield  {journal} {\bibinfo  {journal} {Living
  Reviews in Relativity}\ }\textbf {\bibinfo {volume} {6}},\ \href
  {https://doi.org/10.12942/lrr-2003-5} {10.12942/lrr-2003-5} (\bibinfo {year}
  {2003})\BibitemShut {NoStop}%
\bibitem [{\citenamefont
  {Freire}(2022)}]{https://doi.org/10.48550/arxiv.2204.13468}%
  \BibitemOpen
  \bibfield  {author} {\bibinfo {author} {\bibfnamefont {P.~C.~C.}\
  \bibnamefont {Freire}},\ }\href {https://doi.org/10.48550/ARXIV.2204.13468}
  {\bibinfo {title} {Tests of gravity theories with pulsar timing}} (\bibinfo
  {year} {2022})\BibitemShut {NoStop}%
\bibitem [{\citenamefont {Liu}\ \emph {et~al.}(2012)\citenamefont {Liu},
  \citenamefont {Wex}, \citenamefont {Kramer}, \citenamefont {Cordes},\ and\
  \citenamefont {Lazio}}]{Liu_2012}%
  \BibitemOpen
  \bibfield  {author} {\bibinfo {author} {\bibfnamefont {K.}~\bibnamefont
  {Liu}}, \bibinfo {author} {\bibfnamefont {N.}~\bibnamefont {Wex}}, \bibinfo
  {author} {\bibfnamefont {M.}~\bibnamefont {Kramer}}, \bibinfo {author}
  {\bibfnamefont {J.~M.}\ \bibnamefont {Cordes}},\ and\ \bibinfo {author}
  {\bibfnamefont {T.~J.~W.}\ \bibnamefont {Lazio}},\ }\bibfield  {title}
  {\bibinfo {title} {{PROSPECTS} {FOR} {PROBING} {THE} {SPACETIME} {OF} {WITH}
  {PULSARS}},\ }\href {https://doi.org/10.1088/0004-637x/747/1/1} {\bibfield
  {journal} {\bibinfo  {journal} {The Astrophysical Journal}\ }\textbf
  {\bibinfo {volume} {747}},\ \bibinfo {pages} {1} (\bibinfo {year}
  {2012})}\BibitemShut {NoStop}%
\bibitem [{\citenamefont {Izmailov}\ \emph {et~al.}(2019)\citenamefont
  {Izmailov}, \citenamefont {Zhdanov}, \citenamefont {Bhadra},\ and\
  \citenamefont {Nandi}}]{Izmailov:2019cqr}%
  \BibitemOpen
  \bibfield  {author} {\bibinfo {author} {\bibfnamefont {R.~N.}\ \bibnamefont
  {Izmailov}}, \bibinfo {author} {\bibfnamefont {E.~R.}\ \bibnamefont
  {Zhdanov}}, \bibinfo {author} {\bibfnamefont {A.}~\bibnamefont {Bhadra}},\
  and\ \bibinfo {author} {\bibfnamefont {K.~K.}\ \bibnamefont {Nandi}},\
  }\bibfield  {title} {\bibinfo {title} {{Relative time delay in a spinning
  black hole as a diagnostic for no-hair theorem}},\ }\href
  {https://doi.org/10.1140/epjc/s10052-019-6618-6} {\bibfield  {journal}
  {\bibinfo  {journal} {Eur. Phys. J. C}\ }\textbf {\bibinfo {volume} {79}},\
  \bibinfo {pages} {105} (\bibinfo {year} {2019})}\BibitemShut {NoStop}%
\bibitem [{\citenamefont {{Detweiler}}(1979)}]{detweiler}%
  \BibitemOpen
  \bibfield  {author} {\bibinfo {author} {\bibfnamefont {S.}~\bibnamefont
  {{Detweiler}}},\ }\bibfield  {title} {\bibinfo {title} {{Pulsar timing
  measurements and the search for gravitational waves}},\ }\href
  {https://doi.org/10.1086/157593} {\bibfield  {journal} {\bibinfo  {journal}
  {\apj}\ }\textbf {\bibinfo {volume} {234}},\ \bibinfo {pages} {1100}
  (\bibinfo {year} {1979})}\BibitemShut {NoStop}%
\bibitem [{\citenamefont {Goncharov}\ \emph {et~al.}(2021)\citenamefont
  {Goncharov}, \citenamefont {Shannon}, \citenamefont {Reardon}, \citenamefont
  {Hobbs}, \citenamefont {Zic}, \citenamefont {Bailes}, \citenamefont
  {Cury{\l}o}, \citenamefont {Dai}, \citenamefont {Kerr}, \citenamefont
  {Lower}, \citenamefont {Manchester}, \citenamefont {Mandow}, \citenamefont
  {Middleton}, \citenamefont {Miles}, \citenamefont {Parthasarathy},
  \citenamefont {Thrane}, \citenamefont {Thyagarajan}, \citenamefont {Xue},
  \citenamefont {Zhu}, \citenamefont {Cameron}, \citenamefont {Feng},
  \citenamefont {Luo}, \citenamefont {Russell}, \citenamefont {Sarkissian},
  \citenamefont {Spiewak}, \citenamefont {Wang}, \citenamefont {Wang},
  \citenamefont {Zhang},\ and\ \citenamefont {Zhang}}]{Goncharov_2021}%
  \BibitemOpen
  \bibfield  {author} {\bibinfo {author} {\bibfnamefont {B.}~\bibnamefont
  {Goncharov}}, \bibinfo {author} {\bibfnamefont {R.~M.}\ \bibnamefont
  {Shannon}}, \bibinfo {author} {\bibfnamefont {D.~J.}\ \bibnamefont
  {Reardon}}, \bibinfo {author} {\bibfnamefont {G.}~\bibnamefont {Hobbs}},
  \bibinfo {author} {\bibfnamefont {A.}~\bibnamefont {Zic}}, \bibinfo {author}
  {\bibfnamefont {M.}~\bibnamefont {Bailes}}, \bibinfo {author} {\bibfnamefont
  {M.}~\bibnamefont {Cury{\l}o}}, \bibinfo {author} {\bibfnamefont
  {S.}~\bibnamefont {Dai}}, \bibinfo {author} {\bibfnamefont {M.}~\bibnamefont
  {Kerr}}, \bibinfo {author} {\bibfnamefont {M.~E.}\ \bibnamefont {Lower}},
  \bibinfo {author} {\bibfnamefont {R.~N.}\ \bibnamefont {Manchester}},
  \bibinfo {author} {\bibfnamefont {R.}~\bibnamefont {Mandow}}, \bibinfo
  {author} {\bibfnamefont {H.}~\bibnamefont {Middleton}}, \bibinfo {author}
  {\bibfnamefont {M.~T.}\ \bibnamefont {Miles}}, \bibinfo {author}
  {\bibfnamefont {A.}~\bibnamefont {Parthasarathy}}, \bibinfo {author}
  {\bibfnamefont {E.}~\bibnamefont {Thrane}}, \bibinfo {author} {\bibfnamefont
  {N.}~\bibnamefont {Thyagarajan}}, \bibinfo {author} {\bibfnamefont
  {X.}~\bibnamefont {Xue}}, \bibinfo {author} {\bibfnamefont {X.-J.}\
  \bibnamefont {Zhu}}, \bibinfo {author} {\bibfnamefont {A.~D.}\ \bibnamefont
  {Cameron}}, \bibinfo {author} {\bibfnamefont {Y.}~\bibnamefont {Feng}},
  \bibinfo {author} {\bibfnamefont {R.}~\bibnamefont {Luo}}, \bibinfo {author}
  {\bibfnamefont {C.~J.}\ \bibnamefont {Russell}}, \bibinfo {author}
  {\bibfnamefont {J.}~\bibnamefont {Sarkissian}}, \bibinfo {author}
  {\bibfnamefont {R.}~\bibnamefont {Spiewak}}, \bibinfo {author} {\bibfnamefont
  {S.}~\bibnamefont {Wang}}, \bibinfo {author} {\bibfnamefont {J.~B.}\
  \bibnamefont {Wang}}, \bibinfo {author} {\bibfnamefont {L.}~\bibnamefont
  {Zhang}},\ and\ \bibinfo {author} {\bibfnamefont {S.}~\bibnamefont {Zhang}},\
  }\bibfield  {title} {\bibinfo {title} {On the evidence for a common-spectrum
  process in the search for the nanohertz gravitational-wave background with
  the parkes pulsar timing array},\ }\href
  {https://doi.org/10.3847/2041-8213/ac17f4} {\bibfield  {journal} {\bibinfo
  {journal} {The Astrophysical Journal Letters}\ }\textbf {\bibinfo {volume}
  {917}},\ \bibinfo {pages} {L19} (\bibinfo {year} {2021})}\BibitemShut
  {NoStop}%
\bibitem [{\citenamefont {Chen}\ \emph {et~al.}(2021)\citenamefont {Chen},
  \citenamefont {Caballero}, \citenamefont {Guo}, \citenamefont {Chalumeau},
  \citenamefont {Liu}, \citenamefont {Shaifullah}, \citenamefont {Lee},
  \citenamefont {Babak}, \citenamefont {Desvignes}, \citenamefont
  {Parthasarathy}, \citenamefont {Hu}, \citenamefont {van~der Wateren},
  \citenamefont {Antoniadis}, \citenamefont {Nielsen}, \citenamefont {Bassa},
  \citenamefont {Berthereau}, \citenamefont {Burgay}, \citenamefont {Champion},
  \citenamefont {Cognard}, \citenamefont {Falxa}, \citenamefont {Ferdman},
  \citenamefont {Freire}, \citenamefont {Gair}, \citenamefont {Graikou},
  \citenamefont {Guillemot}, \citenamefont {Jang}, \citenamefont {Janssen},
  \citenamefont {Karuppusamy}, \citenamefont {Keith}, \citenamefont {Kramer},
  \citenamefont {Liu}, \citenamefont {Lyne}, \citenamefont {Main},
  \citenamefont {McKee}, \citenamefont {Mickaliger}, \citenamefont {Perera},
  \citenamefont {Perrodin}, \citenamefont {Petiteau}, \citenamefont {Porayko},
  \citenamefont {Possenti}, \citenamefont {Samajdar}, \citenamefont {Sanidas},
  \citenamefont {Sesana}, \citenamefont {Speri}, \citenamefont {Stappers},
  \citenamefont {Theureau}, \citenamefont {Tiburzi}, \citenamefont {Vecchio},
  \citenamefont {Verbiest}, \citenamefont {Wang}, \citenamefont {Wang},\ and\
  \citenamefont {Xu}}]{ipta}%
  \BibitemOpen
  \bibfield  {author} {\bibinfo {author} {\bibfnamefont {S.}~\bibnamefont
  {Chen}}, \bibinfo {author} {\bibfnamefont {R.~N.}\ \bibnamefont {Caballero}},
  \bibinfo {author} {\bibfnamefont {Y.~J.}\ \bibnamefont {Guo}}, \bibinfo
  {author} {\bibfnamefont {A.}~\bibnamefont {Chalumeau}}, \bibinfo {author}
  {\bibfnamefont {K.}~\bibnamefont {Liu}}, \bibinfo {author} {\bibfnamefont
  {G.}~\bibnamefont {Shaifullah}}, \bibinfo {author} {\bibfnamefont {K.~J.}\
  \bibnamefont {Lee}}, \bibinfo {author} {\bibfnamefont {S.}~\bibnamefont
  {Babak}}, \bibinfo {author} {\bibfnamefont {G.}~\bibnamefont {Desvignes}},
  \bibinfo {author} {\bibfnamefont {A.}~\bibnamefont {Parthasarathy}}, \bibinfo
  {author} {\bibfnamefont {H.}~\bibnamefont {Hu}}, \bibinfo {author}
  {\bibfnamefont {E.}~\bibnamefont {van~der Wateren}}, \bibinfo {author}
  {\bibfnamefont {J.}~\bibnamefont {Antoniadis}}, \bibinfo {author}
  {\bibfnamefont {A.-S.~B.}\ \bibnamefont {Nielsen}}, \bibinfo {author}
  {\bibfnamefont {C.~G.}\ \bibnamefont {Bassa}}, \bibinfo {author}
  {\bibfnamefont {A.}~\bibnamefont {Berthereau}}, \bibinfo {author}
  {\bibfnamefont {M.}~\bibnamefont {Burgay}}, \bibinfo {author} {\bibfnamefont
  {D.~J.}\ \bibnamefont {Champion}}, \bibinfo {author} {\bibfnamefont
  {I.}~\bibnamefont {Cognard}}, \bibinfo {author} {\bibfnamefont
  {M.}~\bibnamefont {Falxa}}, \bibinfo {author} {\bibfnamefont {R.~D.}\
  \bibnamefont {Ferdman}}, \bibinfo {author} {\bibfnamefont {P.~C.~C.}\
  \bibnamefont {Freire}}, \bibinfo {author} {\bibfnamefont {J.~R.}\
  \bibnamefont {Gair}}, \bibinfo {author} {\bibfnamefont {E.}~\bibnamefont
  {Graikou}}, \bibinfo {author} {\bibfnamefont {L.}~\bibnamefont {Guillemot}},
  \bibinfo {author} {\bibfnamefont {J.}~\bibnamefont {Jang}}, \bibinfo {author}
  {\bibfnamefont {G.~H.}\ \bibnamefont {Janssen}}, \bibinfo {author}
  {\bibfnamefont {R.}~\bibnamefont {Karuppusamy}}, \bibinfo {author}
  {\bibfnamefont {M.~J.}\ \bibnamefont {Keith}}, \bibinfo {author}
  {\bibfnamefont {M.}~\bibnamefont {Kramer}}, \bibinfo {author} {\bibfnamefont
  {X.~J.}\ \bibnamefont {Liu}}, \bibinfo {author} {\bibfnamefont {A.~G.}\
  \bibnamefont {Lyne}}, \bibinfo {author} {\bibfnamefont {R.~A.}\ \bibnamefont
  {Main}}, \bibinfo {author} {\bibfnamefont {J.~W.}\ \bibnamefont {McKee}},
  \bibinfo {author} {\bibfnamefont {M.~B.}\ \bibnamefont {Mickaliger}},
  \bibinfo {author} {\bibfnamefont {B.~B.~P.}\ \bibnamefont {Perera}}, \bibinfo
  {author} {\bibfnamefont {D.}~\bibnamefont {Perrodin}}, \bibinfo {author}
  {\bibfnamefont {A.}~\bibnamefont {Petiteau}}, \bibinfo {author}
  {\bibfnamefont {N.~K.}\ \bibnamefont {Porayko}}, \bibinfo {author}
  {\bibfnamefont {A.}~\bibnamefont {Possenti}}, \bibinfo {author}
  {\bibfnamefont {A.}~\bibnamefont {Samajdar}}, \bibinfo {author}
  {\bibfnamefont {S.~A.}\ \bibnamefont {Sanidas}}, \bibinfo {author}
  {\bibfnamefont {A.}~\bibnamefont {Sesana}}, \bibinfo {author} {\bibfnamefont
  {L.}~\bibnamefont {Speri}}, \bibinfo {author} {\bibfnamefont {B.~W.}\
  \bibnamefont {Stappers}}, \bibinfo {author} {\bibfnamefont {G.}~\bibnamefont
  {Theureau}}, \bibinfo {author} {\bibfnamefont {C.}~\bibnamefont {Tiburzi}},
  \bibinfo {author} {\bibfnamefont {A.}~\bibnamefont {Vecchio}}, \bibinfo
  {author} {\bibfnamefont {J.~P.~W.}\ \bibnamefont {Verbiest}}, \bibinfo
  {author} {\bibfnamefont {J.}~\bibnamefont {Wang}}, \bibinfo {author}
  {\bibfnamefont {L.}~\bibnamefont {Wang}},\ and\ \bibinfo {author}
  {\bibfnamefont {H.}~\bibnamefont {Xu}},\ }\bibfield  {title} {\bibinfo
  {title} {Common-red-signal analysis with 24-yr high-precision timing of the
  european pulsar timing array: inferences in the stochastic gravitational-wave
  background search},\ }\href {https://doi.org/10.1093/mnras/stab2833}
  {\bibfield  {journal} {\bibinfo  {journal} {Monthly Notices of the Royal
  Astronomical Society}\ }\textbf {\bibinfo {volume} {508}},\ \bibinfo {pages}
  {4970} (\bibinfo {year} {2021})}\BibitemShut {NoStop}%
\bibitem [{\citenamefont {{Damour}}\ and\ \citenamefont
  {{Deruelle}}(1986)}]{deruelle}%
  \BibitemOpen
  \bibfield  {author} {\bibinfo {author} {\bibfnamefont {T.}~\bibnamefont
  {{Damour}}}\ and\ \bibinfo {author} {\bibfnamefont {N.}~\bibnamefont
  {{Deruelle}}},\ }\bibfield  {title} {\bibinfo {title} {{General relativistic
  celestial mechanics of binary systems. II. The post-Newtonian timing
  formula.}},\ }\href@noop {} {\bibfield  {journal} {\bibinfo  {journal} {Ann.
  Inst. Henri Poincar{\'e} Phys. Th{\'e}or}\ }\textbf {\bibinfo {volume}
  {44}},\ \bibinfo {pages} {263} (\bibinfo {year} {1986})}\BibitemShut
  {NoStop}%
\bibitem [{\citenamefont {{Edwards}}\ \emph {et~al.}(2006)\citenamefont
  {{Edwards}}, \citenamefont {{Hobbs}},\ and\ \citenamefont
  {{Manchester}}}]{2006MNRAS.372.1549E}%
  \BibitemOpen
  \bibfield  {author} {\bibinfo {author} {\bibfnamefont {R.~T.}\ \bibnamefont
  {{Edwards}}}, \bibinfo {author} {\bibfnamefont {G.~B.}\ \bibnamefont
  {{Hobbs}}},\ and\ \bibinfo {author} {\bibfnamefont {R.~N.}\ \bibnamefont
  {{Manchester}}},\ }\bibfield  {title} {\bibinfo {title} {{TEMPO2, a new
  pulsar timing package - II. The timing model and precision estimates}},\
  }\href {https://doi.org/10.1111/j.1365-2966.2006.10870.x} {\bibfield
  {journal} {\bibinfo  {journal} {mnras}\ }\textbf {\bibinfo {volume} {372}},\
  \bibinfo {pages} {1549} (\bibinfo {year} {2006})},\ \Eprint
  {https://arxiv.org/abs/astro-ph/0607664} {arXiv:astro-ph/0607664 [astro-ph]}
  \BibitemShut {NoStop}%
\bibitem [{\citenamefont {Manchester}(2015)}]{Manchester_2015}%
  \BibitemOpen
  \bibfield  {author} {\bibinfo {author} {\bibfnamefont {R.~N.}\ \bibnamefont
  {Manchester}},\ }\bibfield  {title} {\bibinfo {title} {Pulsars and gravity},\
  }\href {https://doi.org/10.1142/s0218271815300189} {\bibfield  {journal}
  {\bibinfo  {journal} {International Journal of Modern Physics D}\ }\textbf
  {\bibinfo {volume} {24}},\ \bibinfo {pages} {1530018} (\bibinfo {year}
  {2015})}\BibitemShut {NoStop}%
\bibitem [{\citenamefont {Ben-Salem}\ and\ \citenamefont
  {Hackmann}(2022)}]{Bilel}%
  \BibitemOpen
  \bibfield  {author} {\bibinfo {author} {\bibfnamefont {B.}~\bibnamefont
  {Ben-Salem}}\ and\ \bibinfo {author} {\bibfnamefont {E.}~\bibnamefont
  {Hackmann}},\ }\bibfield  {title} {\bibinfo {title} {Propagation time delay
  and frame dragging effects of lightlike geodesics in the timing of a pulsar
  orbiting {SgrA}{*}},\ }\href {https://doi.org/10.1093/mnras/stac2337}
  {\bibfield  {journal} {\bibinfo  {journal} {Monthly Notices of the Royal
  Astronomical Society}\ }\textbf {\bibinfo {volume} {516}},\ \bibinfo {pages}
  {1768} (\bibinfo {year} {2022})}\BibitemShut {NoStop}%
\bibitem [{\citenamefont {Hackmann}\ and\ \citenamefont
  {Dhani}(2019)}]{Eva_2019}%
  \BibitemOpen
  \bibfield  {author} {\bibinfo {author} {\bibfnamefont {E.}~\bibnamefont
  {Hackmann}}\ and\ \bibinfo {author} {\bibfnamefont {A.}~\bibnamefont
  {Dhani}},\ }\bibfield  {title} {\bibinfo {title} {The propagation delay in
  the timing of a pulsar orbiting a supermassive black hole},\ }\bibfield
  {journal} {\bibinfo  {journal} {General Relativity and Gravitation}\ }\textbf
  {\bibinfo {volume} {51}},\ \href {https://doi.org/10.1007/s10714-019-2517-2}
  {10.1007/s10714-019-2517-2} (\bibinfo {year} {2019})\BibitemShut {NoStop}%
\bibitem [{\citenamefont {Zhang}\ and\ \citenamefont
  {Saha}(2017)}]{Zhang_2017}%
  \BibitemOpen
  \bibfield  {author} {\bibinfo {author} {\bibfnamefont {F.}~\bibnamefont
  {Zhang}}\ and\ \bibinfo {author} {\bibfnamefont {P.}~\bibnamefont {Saha}},\
  }\bibfield  {title} {\bibinfo {title} {Probing the spinning of the massive
  black hole in the galactic center via pulsar timing: A full relativistic
  treatment},\ }\href {https://doi.org/10.3847/1538-4357/aa8f47} {\bibfield
  {journal} {\bibinfo  {journal} {The Astrophysical Journal}\ }\textbf
  {\bibinfo {volume} {849}},\ \bibinfo {pages} {33} (\bibinfo {year}
  {2017})}\BibitemShut {NoStop}%
\bibitem [{\citenamefont {Kimpson}\ \emph {et~al.}(2019)\citenamefont
  {Kimpson}, \citenamefont {Wu},\ and\ \citenamefont {Zane}}]{Kimpson_2019}%
  \BibitemOpen
  \bibfield  {author} {\bibinfo {author} {\bibfnamefont {T.}~\bibnamefont
  {Kimpson}}, \bibinfo {author} {\bibfnamefont {K.}~\bibnamefont {Wu}},\ and\
  \bibinfo {author} {\bibfnamefont {S.}~\bibnamefont {Zane}},\ }\bibfield
  {title} {\bibinfo {title} {Pulsar timing in extreme mass ratio binaries: a
  general relativistic approach},\ }\href
  {https://doi.org/10.1093/mnras/stz845} {\bibfield  {journal} {\bibinfo
  {journal} {Monthly Notices of the Royal Astronomical Society}\ }\textbf
  {\bibinfo {volume} {486}},\ \bibinfo {pages} {360} (\bibinfo {year}
  {2019})}\BibitemShut {NoStop}%
\bibitem [{\citenamefont {Farrah}\ \emph
  {et~al.}(2023{\natexlab{a}})\citenamefont {Farrah}, \citenamefont {Petty},
  \citenamefont {Croker}, \citenamefont {Tarl{\'{e} }}, \citenamefont {Zevin},
  \citenamefont {Hatziminaoglou}, \citenamefont {Shankar}, \citenamefont
  {Wang}, \citenamefont {Clements}, \citenamefont {Efstathiou}, \citenamefont
  {Lacy}, \citenamefont {Nishimura}, \citenamefont {Afonso}, \citenamefont
  {Pearson},\ and\ \citenamefont {Pitchford}}]{BH_DE_inside}%
  \BibitemOpen
  \bibfield  {author} {\bibinfo {author} {\bibfnamefont {D.}~\bibnamefont
  {Farrah}}, \bibinfo {author} {\bibfnamefont {S.}~\bibnamefont {Petty}},
  \bibinfo {author} {\bibfnamefont {K.~S.}\ \bibnamefont {Croker}}, \bibinfo
  {author} {\bibfnamefont {G.}~\bibnamefont {Tarl{\'{e} }}}, \bibinfo {author}
  {\bibfnamefont {M.}~\bibnamefont {Zevin}}, \bibinfo {author} {\bibfnamefont
  {E.}~\bibnamefont {Hatziminaoglou}}, \bibinfo {author} {\bibfnamefont
  {F.}~\bibnamefont {Shankar}}, \bibinfo {author} {\bibfnamefont
  {L.}~\bibnamefont {Wang}}, \bibinfo {author} {\bibfnamefont {D.~L.}\
  \bibnamefont {Clements}}, \bibinfo {author} {\bibfnamefont {A.}~\bibnamefont
  {Efstathiou}}, \bibinfo {author} {\bibfnamefont {M.}~\bibnamefont {Lacy}},
  \bibinfo {author} {\bibfnamefont {K.~A.}\ \bibnamefont {Nishimura}}, \bibinfo
  {author} {\bibfnamefont {J.}~\bibnamefont {Afonso}}, \bibinfo {author}
  {\bibfnamefont {C.}~\bibnamefont {Pearson}},\ and\ \bibinfo {author}
  {\bibfnamefont {L.~K.}\ \bibnamefont {Pitchford}},\ }\bibfield  {title}
  {\bibinfo {title} {A preferential growth channel for supermassive black holes
  in elliptical galaxies at z $\lesssim$ 2},\ }\href
  {https://doi.org/10.3847/1538-4357/acac2e} {\bibfield  {journal} {\bibinfo
  {journal} {The Astrophysical Journal}\ }\textbf {\bibinfo {volume} {943}},\
  \bibinfo {pages} {133} (\bibinfo {year} {2023}{\natexlab{a}})}\BibitemShut
  {NoStop}%
\bibitem [{\citenamefont {Farrah}\ \emph
  {et~al.}(2023{\natexlab{b}})\citenamefont {Farrah}, \citenamefont {Croker},
  \citenamefont {Zevin}, \citenamefont {Tarl{\'{e} }}, \citenamefont {Faraoni},
  \citenamefont {Petty}, \citenamefont {Afonso}, \citenamefont {Fernandez},
  \citenamefont {Nishimura}, \citenamefont {Pearson}, \citenamefont {Wang},
  \citenamefont {Clements}, \citenamefont {Efstathiou}, \citenamefont
  {Hatziminaoglou}, \citenamefont {Lacy}, \citenamefont {McPartland},
  \citenamefont {Pitchford}, \citenamefont {Sakai},\ and\ \citenamefont
  {Weiner}}]{BH_DE_inside_2}%
  \BibitemOpen
  \bibfield  {author} {\bibinfo {author} {\bibfnamefont {D.}~\bibnamefont
  {Farrah}}, \bibinfo {author} {\bibfnamefont {K.~S.}\ \bibnamefont {Croker}},
  \bibinfo {author} {\bibfnamefont {M.}~\bibnamefont {Zevin}}, \bibinfo
  {author} {\bibfnamefont {G.}~\bibnamefont {Tarl{\'{e} }}}, \bibinfo {author}
  {\bibfnamefont {V.}~\bibnamefont {Faraoni}}, \bibinfo {author} {\bibfnamefont
  {S.}~\bibnamefont {Petty}}, \bibinfo {author} {\bibfnamefont
  {J.}~\bibnamefont {Afonso}}, \bibinfo {author} {\bibfnamefont
  {N.}~\bibnamefont {Fernandez}}, \bibinfo {author} {\bibfnamefont {K.~A.}\
  \bibnamefont {Nishimura}}, \bibinfo {author} {\bibfnamefont {C.}~\bibnamefont
  {Pearson}}, \bibinfo {author} {\bibfnamefont {L.}~\bibnamefont {Wang}},
  \bibinfo {author} {\bibfnamefont {D.~L.}\ \bibnamefont {Clements}}, \bibinfo
  {author} {\bibfnamefont {A.}~\bibnamefont {Efstathiou}}, \bibinfo {author}
  {\bibfnamefont {E.}~\bibnamefont {Hatziminaoglou}}, \bibinfo {author}
  {\bibfnamefont {M.}~\bibnamefont {Lacy}}, \bibinfo {author} {\bibfnamefont
  {C.}~\bibnamefont {McPartland}}, \bibinfo {author} {\bibfnamefont {L.~K.}\
  \bibnamefont {Pitchford}}, \bibinfo {author} {\bibfnamefont {N.}~\bibnamefont
  {Sakai}},\ and\ \bibinfo {author} {\bibfnamefont {J.}~\bibnamefont
  {Weiner}},\ }\bibfield  {title} {\bibinfo {title} {Observational evidence for
  cosmological coupling of black holes and its implications for an
  astrophysical source of dark energy},\ }\href
  {https://doi.org/10.3847/2041-8213/acb704} {\bibfield  {journal} {\bibinfo
  {journal} {The Astrophysical Journal Letters}\ }\textbf {\bibinfo {volume}
  {944}},\ \bibinfo {pages} {L31} (\bibinfo {year}
  {2023}{\natexlab{b}})}\BibitemShut {NoStop}%
\bibitem [{\citenamefont {Kiselev}(2003)}]{Kiselev:2003}%
  \BibitemOpen
  \bibfield  {author} {\bibinfo {author} {\bibfnamefont {V.~V.}\ \bibnamefont
  {Kiselev}},\ }\bibfield  {title} {\bibinfo {title} {{Quintessential solution
  of dark matter rotation curves and its simulation by extra dimensions}},\
  }\href@noop {} {\  (\bibinfo {year} {2003})},\ \Eprint
  {https://arxiv.org/abs/gr-qc/0303031} {arXiv:gr-qc/0303031} \BibitemShut
  {NoStop}%
\bibitem [{\citenamefont {Ghosh}(2016)}]{Ghosh}%
  \BibitemOpen
  \bibfield  {author} {\bibinfo {author} {\bibfnamefont {S.~G.}\ \bibnamefont
  {Ghosh}},\ }\bibfield  {title} {\bibinfo {title} {{Rotating black hole and
  quintessence}},\ }\href {https://doi.org/10.1140/epjc/s10052-016-4051-7}
  {\bibfield  {journal} {\bibinfo  {journal} {Eur. Phys. J. C}\ }\textbf
  {\bibinfo {volume} {76}},\ \bibinfo {pages} {222} (\bibinfo {year} {2016})},\
  \Eprint {https://arxiv.org/abs/1512.05476} {arXiv:1512.05476 [gr-qc]}
  \BibitemShut {NoStop}%
\bibitem [{\citenamefont {Toshmatov}\ \emph {et~al.}(2017)\citenamefont
  {Toshmatov}, \citenamefont {Stuchl\'\i{}k},\ and\ \citenamefont
  {Ahmedov}}]{Toshmatov:2015npp}%
  \BibitemOpen
  \bibfield  {author} {\bibinfo {author} {\bibfnamefont {B.}~\bibnamefont
  {Toshmatov}}, \bibinfo {author} {\bibfnamefont {Z.}~\bibnamefont
  {Stuchl\'\i{}k}},\ and\ \bibinfo {author} {\bibfnamefont {B.}~\bibnamefont
  {Ahmedov}},\ }\bibfield  {title} {\bibinfo {title} {{Rotating black hole
  solutions with quintessential energy}},\ }\href
  {https://doi.org/10.1140/epjp/i2017-11373-4} {\bibfield  {journal} {\bibinfo
  {journal} {Eur. Phys. J. Plus}\ }\textbf {\bibinfo {volume} {132}},\ \bibinfo
  {pages} {98} (\bibinfo {year} {2017})},\ \Eprint
  {https://arxiv.org/abs/1512.01498} {arXiv:1512.01498 [gr-qc]} \BibitemShut
  {NoStop}%
\bibitem [{\citenamefont {Melia}(2015)}]{Melia:2014vva}%
  \BibitemOpen
  \bibfield  {author} {\bibinfo {author} {\bibfnamefont {F.}~\bibnamefont
  {Melia}},\ }\bibfield  {title} {\bibinfo {title} {{The Cosmic Equation of
  State}},\ }\href {https://doi.org/10.1007/s10509-014-2211-5} {\bibfield
  {journal} {\bibinfo  {journal} {Astrophys. Space Sci.}\ }\textbf {\bibinfo
  {volume} {356}},\ \bibinfo {pages} {393} (\bibinfo {year} {2015})},\ \Eprint
  {https://arxiv.org/abs/1411.5771} {arXiv:1411.5771 [astro-ph.CO]}
  \BibitemShut {NoStop}%
\bibitem [{\citenamefont {Carleo}\ \emph {et~al.}(2022)\citenamefont {Carleo},
  \citenamefont {Lambiase},\ and\ \citenamefont {Mastrototaro}}]{carleo1}%
  \BibitemOpen
  \bibfield  {author} {\bibinfo {author} {\bibfnamefont {A.}~\bibnamefont
  {Carleo}}, \bibinfo {author} {\bibfnamefont {G.}~\bibnamefont {Lambiase}},\
  and\ \bibinfo {author} {\bibfnamefont {L.}~\bibnamefont {Mastrototaro}},\
  }\bibfield  {title} {\bibinfo {title} {{Energy extraction via magnetic
  reconnection in Lorentz breaking Kerr\textendash{}Sen and Kiselev black
  holes}},\ }\href {https://doi.org/10.1140/epjc/s10052-022-10751-w} {\bibfield
   {journal} {\bibinfo  {journal} {Eur. Phys. J. C}\ }\textbf {\bibinfo
  {volume} {82}},\ \bibinfo {pages} {776} (\bibinfo {year} {2022})},\ \Eprint
  {https://arxiv.org/abs/2206.12988} {arXiv:2206.12988 [gr-qc]} \BibitemShut
  {NoStop}%
\bibitem [{\citenamefont {Akiyama}\ \emph {et~al.}(2019)\citenamefont {Akiyama}
  \emph {et~al.}}]{M87}%
  \BibitemOpen
  \bibfield  {author} {\bibinfo {author} {\bibfnamefont {K.}~\bibnamefont
  {Akiyama}} \emph {et~al.} (\bibinfo {collaboration} {Event Horizon
  Telescope}),\ }\bibfield  {title} {\bibinfo {title} {{First M87 Event Horizon
  Telescope Results. I. The Shadow of the Supermassive Black Hole}},\ }\href
  {https://doi.org/10.3847/2041-8213/ab0ec7} {\bibfield  {journal} {\bibinfo
  {journal} {Astrophys. J. Lett.}\ }\textbf {\bibinfo {volume} {875}},\
  \bibinfo {pages} {L1} (\bibinfo {year} {2019})},\ \Eprint
  {https://arxiv.org/abs/1906.11238} {arXiv:1906.11238 [astro-ph.GA]}
  \BibitemShut {NoStop}%
\bibitem [{\citenamefont {{Pratap Singh}}(2017)}]{2017}%
  \BibitemOpen
  \bibfield  {author} {\bibinfo {author} {\bibfnamefont {B.}~\bibnamefont
  {{Pratap Singh}}},\ }\bibfield  {title} {\bibinfo {title} {{Rotating charged
  black holes shadow in quintessence}},\ }\href@noop {} {\bibfield  {journal}
  {\bibinfo  {journal} {arXiv e-prints}\ ,\ \bibinfo {eid} {arXiv:1711.02898}}
  (\bibinfo {year} {2017})},\ \Eprint {https://arxiv.org/abs/1711.02898}
  {arXiv:1711.02898 [gr-qc]} \BibitemShut {NoStop}%
\bibitem [{\citenamefont {Carter}(1968)}]{carter}%
  \BibitemOpen
  \bibfield  {author} {\bibinfo {author} {\bibfnamefont {B.}~\bibnamefont
  {Carter}},\ }\bibfield  {title} {\bibinfo {title} {Global structure of the
  kerr family of gravitational fields},\ }\href
  {https://doi.org/10.1103/PhysRev.174.1559} {\bibfield  {journal} {\bibinfo
  {journal} {Phys. Rev.}\ }\textbf {\bibinfo {volume} {174}},\ \bibinfo {pages}
  {1559} (\bibinfo {year} {1968})}\BibitemShut {NoStop}%
\bibitem [{\citenamefont {Gralla}\ and\ \citenamefont
  {Lupsasca}(2020)}]{Gralla_2020}%
  \BibitemOpen
  \bibfield  {author} {\bibinfo {author} {\bibfnamefont {S.~E.}\ \bibnamefont
  {Gralla}}\ and\ \bibinfo {author} {\bibfnamefont {A.}~\bibnamefont
  {Lupsasca}},\ }\bibfield  {title} {\bibinfo {title} {Null geodesics of the
  kerr exterior},\ }\bibfield  {journal} {\bibinfo  {journal} {Physical Review
  D}\ }\textbf {\bibinfo {volume} {101}},\ \href
  {https://doi.org/10.1103/physrevd.101.044032} {10.1103/physrevd.101.044032}
  (\bibinfo {year} {2020})\BibitemShut {NoStop}%
\bibitem [{\citenamefont {Dexter}\ and\ \citenamefont
  {Agol}(2009)}]{Dexter_2009}%
  \BibitemOpen
  \bibfield  {author} {\bibinfo {author} {\bibfnamefont {J.}~\bibnamefont
  {Dexter}}\ and\ \bibinfo {author} {\bibfnamefont {E.}~\bibnamefont {Agol}},\
  }\bibfield  {title} {\bibinfo {title} {A {FAST} {NEW} {PUBLIC} {CODE} {FOR}
  {COMPUTING} {PHOTON} {ORBITS} {IN} a {KERR} {SPACETIME}},\ }\href
  {https://doi.org/10.1088/0004-637x/696/2/1616} {\bibfield  {journal}
  {\bibinfo  {journal} {The Astrophysical Journal}\ }\textbf {\bibinfo {volume}
  {696}},\ \bibinfo {pages} {1616} (\bibinfo {year} {2009})}\BibitemShut
  {NoStop}%
\bibitem [{\citenamefont {{Blandford}}\ and\ \citenamefont
  {{Teukolsky}}(1976)}]{Blandford1976}%
  \BibitemOpen
  \bibfield  {author} {\bibinfo {author} {\bibfnamefont {R.}~\bibnamefont
  {{Blandford}}}\ and\ \bibinfo {author} {\bibfnamefont {S.~A.}\ \bibnamefont
  {{Teukolsky}}},\ }\bibfield  {title} {\bibinfo {title} {{Arrival-time
  analysis for a pulsar in a binary system.}},\ }\href
  {https://doi.org/10.1086/154315} {\bibfield  {journal} {\bibinfo  {journal}
  {Astrophys. J.}\ }\textbf {\bibinfo {volume} {205}},\ \bibinfo {pages} {580}
  (\bibinfo {year} {1976})}\BibitemShut {NoStop}%
\bibitem [{\citenamefont {Torne}\ \emph {et~al.}(2021)\citenamefont {Torne},
  \citenamefont {Desvignes}, \citenamefont {Eatough}, \citenamefont {Kramer},
  \citenamefont {Karuppusamy}, \citenamefont {Liu}, \citenamefont {Noutsos},
  \citenamefont {Wharton}, \citenamefont {Kramer}, \citenamefont {Navarro},
  \citenamefont {Paubert}, \citenamefont {Sanchez}, \citenamefont
  {Sanchez-Portal}, \citenamefont {Schuster}, \citenamefont {Falcke},\ and\
  \citenamefont {Rezzolla}}]{new_pulsars}%
  \BibitemOpen
  \bibfield  {author} {\bibinfo {author} {\bibfnamefont {P.}~\bibnamefont
  {Torne}}, \bibinfo {author} {\bibfnamefont {G.}~\bibnamefont {Desvignes}},
  \bibinfo {author} {\bibfnamefont {R.~P.}\ \bibnamefont {Eatough}}, \bibinfo
  {author} {\bibfnamefont {M.}~\bibnamefont {Kramer}}, \bibinfo {author}
  {\bibfnamefont {R.}~\bibnamefont {Karuppusamy}}, \bibinfo {author}
  {\bibfnamefont {K.}~\bibnamefont {Liu}}, \bibinfo {author} {\bibfnamefont
  {A.}~\bibnamefont {Noutsos}}, \bibinfo {author} {\bibfnamefont
  {R.}~\bibnamefont {Wharton}}, \bibinfo {author} {\bibfnamefont
  {C.}~\bibnamefont {Kramer}}, \bibinfo {author} {\bibfnamefont
  {S.}~\bibnamefont {Navarro}}, \bibinfo {author} {\bibfnamefont
  {G.}~\bibnamefont {Paubert}}, \bibinfo {author} {\bibfnamefont
  {S.}~\bibnamefont {Sanchez}}, \bibinfo {author} {\bibfnamefont
  {M.}~\bibnamefont {Sanchez-Portal}}, \bibinfo {author} {\bibfnamefont
  {K.~F.}\ \bibnamefont {Schuster}}, \bibinfo {author} {\bibfnamefont
  {H.}~\bibnamefont {Falcke}},\ and\ \bibinfo {author} {\bibfnamefont
  {L.}~\bibnamefont {Rezzolla}},\ }\bibfield  {title} {\bibinfo {title}
  {Searching for pulsars in the galactic centre at 3 and 2 mm},\ }\href
  {https://doi.org/10.1051/0004-6361/202140775} {\bibfield  {journal} {\bibinfo
   {journal} {Astronomy {\&} Astrophysics}\ }\textbf {\bibinfo {volume}
  {650}},\ \bibinfo {pages} {A95} (\bibinfo {year} {2021})}\BibitemShut
  {NoStop}%
\bibitem [{\citenamefont {{Pfahl}}\ and\ \citenamefont
  {{Loeb}}(2004)}]{probe_2004}%
  \BibitemOpen
  \bibfield  {author} {\bibinfo {author} {\bibfnamefont {E.}~\bibnamefont
  {{Pfahl}}}\ and\ \bibinfo {author} {\bibfnamefont {A.}~\bibnamefont
  {{Loeb}}},\ }\bibfield  {title} {\bibinfo {title} {{Probing the Spacetime
  around Sagittarius A* with Radio Pulsars}},\ }\href
  {https://doi.org/10.1086/423975} {\bibfield  {journal} {\bibinfo  {journal}
  {\apj}\ }\textbf {\bibinfo {volume} {615}},\ \bibinfo {pages} {253} (\bibinfo
  {year} {2004})},\ \Eprint {https://arxiv.org/abs/astro-ph/0309744}
  {arXiv:astro-ph/0309744 [astro-ph]} \BibitemShut {NoStop}%
\bibitem [{\citenamefont {Yusef-Zadeh}\ \emph {et~al.}(2015)\citenamefont
  {Yusef-Zadeh}, \citenamefont {Diesing}, \citenamefont {Wardle}, \citenamefont
  {Sjouwerman}, \citenamefont {Royster}, \citenamefont {Cotton}, \citenamefont
  {Roberts},\ and\ \citenamefont {Heinke}}]{distance_2015}%
  \BibitemOpen
  \bibfield  {author} {\bibinfo {author} {\bibfnamefont {F.}~\bibnamefont
  {Yusef-Zadeh}}, \bibinfo {author} {\bibfnamefont {R.}~\bibnamefont
  {Diesing}}, \bibinfo {author} {\bibfnamefont {M.}~\bibnamefont {Wardle}},
  \bibinfo {author} {\bibfnamefont {L.~O.}\ \bibnamefont {Sjouwerman}},
  \bibinfo {author} {\bibfnamefont {M.}~\bibnamefont {Royster}}, \bibinfo
  {author} {\bibfnamefont {W.~D.}\ \bibnamefont {Cotton}}, \bibinfo {author}
  {\bibfnamefont {D.}~\bibnamefont {Roberts}},\ and\ \bibinfo {author}
  {\bibfnamefont {C.}~\bibnamefont {Heinke}},\ }\bibfield  {title} {\bibinfo
  {title} {{RADIO} {CONTINUUM} {EMISSION} {FROM} {THE} {MAGNETAR} {SGR}
  j1745-2900: {INTERACTION} {WITH} {GAS} {ORBITING} sgr a*},\ }\href
  {https://doi.org/10.1088/2041-8205/811/2/l35} {\bibfield  {journal} {\bibinfo
   {journal} {The Astrophysical Journal}\ }\textbf {\bibinfo {volume} {811}},\
  \bibinfo {pages} {L35} (\bibinfo {year} {2015})}\BibitemShut {NoStop}%
\bibitem [{\citenamefont {Boas~Jr}(1954)}]{higher}%
  \BibitemOpen
  \bibfield  {author} {\bibinfo {author} {\bibfnamefont {R.}~\bibnamefont
  {Boas~Jr}},\ }\bibfield  {title} {\bibinfo {title} {Higher transcendental
  functions, vols. i and ii.},\ }\href@noop {} {\bibfield  {journal} {\bibinfo
  {journal} {Science}\ }\textbf {\bibinfo {volume} {120}},\ \bibinfo {pages}
  {302} (\bibinfo {year} {1954})}\BibitemShut {NoStop}%
\end{thebibliography}
\end{document}